\documentclass[10pt,a4paper]{article}

\usepackage{jheppub}

\usepackage{graphicx}
\usepackage{bm}
\usepackage{amsmath}
\usepackage{amssymb}
\usepackage{amsfonts}
\usepackage{float}
\usepackage{hyperref}
\usepackage{dsfont}  
\usepackage{slashed}  
\usepackage{booktabs}
\usepackage{multirow}
\usepackage{subfigure}
\usepackage[sort&compress]{natbib}
\usepackage{xcolor}
\usepackage{ulem}
\usepackage{mathabx}

\newcommand{\be}{\begin{equation}}  
\newcommand{\ee}{\end{equation}}  
\newcommand{\beq}{\begin{eqnarray}} 
\newcommand{\eeq}{\end{eqnarray}}

\newcommand{\bea}{\begin{eqnarray}}
\newcommand{\eea}{\end{eqnarray}}

\newcommand{\Slash}[1]{{\ooalign{\hfil/\hfil\crcr$#1$}}}
\newcommand{\nn}{\nonumber \\}

\usepackage{braket}
\usepackage{slashed}
\usepackage{cancel}
\newcommand{\w}[1]{\widetilde{#1}}
\newcommand{\f}[1]{\mathcal{#1}}
\newcommand{\sbar}[1]{\slashed{\bar{#1}}}
\newcommand{\s}[1]{\slashed{#1}}

\begin{document}

\title{ Generalized parton distributions and gravitational form factors at large momentum transfer }

\author{Yoshitaka Hatta and Jakob Schoenleber }
\affiliation{Physics Department, Brookhaven National Laboratory, Upton, NY 11973, USA}
\affiliation{RIKEN BNL Research Center, Brookhaven National Laboratory, Upton, NY 11973, USA}

\abstract{

Within the soft collinear effective theory (SCET), we derive a factorization theorem  which resums Sudakov logarithms $(\alpha_s\ln^2(-t))^n$ to all orders in the quark-in-quark generalized parton distribution (GPD) at large momentum transfer $t$, and  perform a  consistency check to one-loop. We  show that the same Sudakov factor  appears in the `Feynman' contribution to the GPDs of the nucleon.  Our result enables the resummation of all the large logarithms $\ln Q^2$ and $\ln^2t$ in  exclusive processes with two hard scales $\Lambda_{\rm QCD}^2\ll |t| \ll Q^2$. We also present a SCET power counting analysis of the Feynman contributions to the GPDs  and show that the $x$-dependence of GPDs factorizes at large-$t$ with controlled corrections. This in particular implies that any ratio of GPD moments such as the electromagnetic and gravitational form factors (GFF)  is perturbatively calculable in this approximation. 
Furthermore, we identify a novel order $\alpha_s$ power-law $t$-dependence in the  GPD  and the $D$-type GFF that  will    dominate over the standard order $(\alpha_s^2)$ `leading twist' asymptotic  contribution  in the phenomenologically relevant region of $t$.  }

\maketitle

\section{Introduction}

The study of the electromagnetic form factor of hadrons $F(t)$  at large momentum transfer $|t|\gg \Lambda_{\rm QCD}^2$ has a long and winding history  \cite{Punjabi:2015bba}. One of the  earliest theoretical ideas,   based on the parton model, is the so-called  Feynman  contribution (also called the soft or overlap  contribution) where one `active' parton absorbs the virtual photon  and recombines with the  `spectator' partons. This simple picture predicted power-law scaling rules $F(t)\sim 1/t^n$ whose exponent $n$ is related to the $x\to 1$ behavior of the structure functions  \cite{Drell:1969km,West:1970av}. Subsequently, an alternative scenario, in which all the valence partons are involved, was put forward \cite{Brodsky:1973kr}. The asymptotic behavior again follows a power-law, but $n$ is now related to the number of constituents in the hadron's lowest Fock component.  With the advent of QCD, the latter approach was promoted to a rigorous collinear  factorization theorem \cite{Chernyak:1977fk,Efremov:1979qk,Lepage:1980fj} and its transverse momentum dependent version   \cite{Li:1992nu,Li:1992ce}. 
Up to power corrections, the  form factor is represented by the convolution of the hard kernel, in which hard gluons are exchanged between all the valence partons, and   nonperturbative distribution amplitudes.  From the perturbative QCD viewpoint, the Feynman contribution is a subleading  correction highly suppressed by the Sudakov effect in the limit $|t|\to \infty$ \cite{Lepage:1980fj}. However, concerns  have been raised, and persist to this day, that in practice the leading perturbative QCD contribution appears to be numerically too small to explain the experimental data \cite{Nesterenko:1983ef,Isgur:1984jm,Isgur:1988iw,Radyushkin:1990te,Jakob:1993iw,Diehl:1998kh,Radyushkin:1998rt,Braun:2006hz}. In the case of the pion form factor, higher order perturbative QCD corrections tend to  ameliorate the discrepancy \cite{Ji:2024iak}. However, the leading twist hard kernel for the proton form factor is suppressed by an extra power of   $\alpha_s/\pi\sim 0.1$ compared to the pion case. It has been argued  that the Feynman contribution, being formally subleading in $1/t$  but not suppressed by $\alpha_s$ \cite{Nesterenko:1983ef,Braun:2006hz,Braun:2001tj}, may actually dominate the form factor in the entire range of $t$ experimentally accessible  in the foreseeable future. Moreover, in the asymptotic region $|t|\to \infty$, there are Feynman-like contributions to the nucleon form factor that have the same power-law behavior in $1/t$ (although suppressed in $\alpha_s$) as the leading twist contribution \cite{Duncan:1979hi,Milshtein:1982js,Kivel:2010ns,Kivel:2012mf,Kivel:2013sya}. 

Since the form factor $F(t)$ is the first moment of the generalized parton distribution (GPD) $F(t)=\int dx F(x,\xi,t)$ \cite{Diehl:2003ny,Belitsky:2005qn}, similar questions arise with regard to the large-$t$ behavior of GPDs.  The  leading twist contribution is given by a convolution between the hard factor, in which all the valence quarks are connected by hard gluons, and the hadron distribution amplitudes  \cite{Diehl:1999ek,Vogt:2001if,Hoodbhoy:2003uu}, similar to the factorization of the form factor. The analog of the Feynman contribution also exists and is called the `overlap representation' of GPDs \cite{Diehl:2000xz}  where one energetic quark absorbs the total momentum transfer $t$. The remaining  partons must be `wee' partons, so that they do not need to be `redirected' by  hard interactions in order to recombine with the struck quark. Given the situation with the form factor described above, we expect that the Feynman contribution is a significant, or even dominant contribution to GPDs for realistic values of $t$. 

In this paper, we first discuss radiative QCD corrections to the `quark-in-quark' GPD (quark GPD of a quark) which we view as a building block of `quark-in-hadron' GPDs in the overlap representation. More precisely, we  resum the Sudakov double logarithms $(\alpha_s\ln^2 t)^n$  in the regime $|t|\gg \Lambda^2_{\rm QCD}$ to all orders in perturbation theory. 
The analogous problem is the Sudakov form factor of a quark $F_q(t) \sim \langle p_2 | \bar \psi(0) \gamma^{\mu} \psi(0) | p_1 \rangle$  (or an electron in QED), which is essentially the local version of the quark-in-quark GPD from our perspective. The formal factorization of the Sudakov form factor $F_q$ with quark external states is well understood and has become a textbook material \cite{Collins:2011zzd, Becher:2014oda}.\footnote{We remark here that only the massless case is well understood and there are still open questions regarding the massive fermion case \cite{Smirnov:1999bza}.} In contrast, in the GPD literature,  Sudakov double logarithms   have been identified  in one-loop calculations in different contexts   \cite{Huang:2001ej,Radyushkin:2021fel,Bhattacharya:2023wvy}, but so far there has been no systematic, all-order treatment.  Using the soft collinear effective theory (SCET) \cite{Bauer:2000yr,Bauer:2001ct, Bauer:2001yt,Bauer:2002nz}, we will show that these logarithms can be resummed and factorized in a way completely analogous to the quark form factor $F_q(t)$.

Once the factorization of the quark-in-quark GPD is established, in Sec.~\ref{proton new} we will integrate it into the quark-in-proton GPD (i.e., ordinary GPD) and discuss the large-$t$ behavior of the GPD. In Sec.~\ref{sec: SCET soft}, we perform the SCET power counting analysis to derive  the $x$ and $t$-dependencies of the GPD to leading power in $1/t$. Moreover, we will  discuss the implications of our results for the large-$t$ behavior of the gravitational form factors (GFFs). Then in Sec.~\ref{lc}, we employ  the overlap representation of GPDs in order to include power corrections as well as possible nonperturbative effects.

Our work is partially motivated by the recent analysis of Deeply Virtual Compton Scattering (DVCS) \cite{Bhattacharya:2023wvy} which  includes the regime $\Lambda^2_{\rm QCD}\ll |t| \ll Q^2$ where $Q^2$ is the photon virtuality. In the usual collinear factorization \cite{Collins:1998be,Ji:1998xh} recently extended to twist-three \cite{Schoenleber:2024ihr}, the large logarithms $\alpha_s\ln Q^2$ arising in the perturbative calculation of the Compton amplitude can be factorized and absorbed into GPDs via the standard renormalization procedure. In this framework, one sets $t=0$ in the partonic amplitudes from the outset,  and the entire  $t$-dependence of GPDs is regarded as  nonperturbative. However,  explicit calculations with $t\neq 0$ have revealed the presence of double (Sudakov) logarithms $\alpha_s \ln^2 t$. At one-loop, they can also be absorbed into GPDs  \cite{Bhattacharya:2023wvy}, but their all-order treatment has not been addressed in the GPD literature.   The present work paves the way for the all-order resummation and factorization of all the large logarithms $(\alpha_s\ln Q^2)^n$ and $(\alpha_s\ln^2 t)^n$ (and subleading logarithms in $t$) in the said kinematical region of DVCS and related processes such as wide-angle Compton scattering \cite{Radyushkin:1998rt,Huang:2001ej}.

\section{Factorization of quark-in-quark GPD in SCET}
\label{quarkin}

\subsection{Quark-in-quark GPD}

Our starting point is the quark GPD of a quark defined by 
\begin{align}
F(x,y,\xi,t, p^2_1 , p^2_2) \bar{u}(p_2)\sbar n u(p_1) = \bar n\cdot P \int_{-\infty}^{\infty} \frac{dz}{2\pi} e^{izx \bar n\cdot P} \widetilde F(z\bar n/2,-z\bar n/2, p_1,p_2),  \label{def} 
\end{align}
where
\beq
\widetilde F(z_1 \bar n,z_2 \bar  n,p_1,p_2) = \bra{p_2} \bar \psi(z_2 \bar n) \sbar n W(z_2 \bar n,z_1 \bar n) \psi(z_1 \bar n) \ket{p_1}. \label{before}
\eeq 
$|p_1\rangle$ and $|p_2\rangle$ are single massless quark external states  which we assume to be slightly off-shell $p^2_1,p_2^2<0$ for the sake of IR regularization.  The variable $x$ (more precisely, $\frac{x\pm \xi}{y\pm \xi}$, see below) represents the momentum fraction of the external quark carried by the `probe' quark. 
$\bar n^{\mu}$ is a fixed light-like reference vector that projects out the large momentum component $\bar n \cdot p_1 \sim \bar n \cdot p_2$. Definitions of GPDs in the literature vary depending on the different definitions of the skewness variable $\xi$, which is usually defined with respect to the specific process that the GPD is part of. Here, we define the average momentum and $\xi$ as
\begin{align}
P^\mu=\frac{p^\mu_1+p_2^\mu}{2y}, \qquad \xi=\frac{\bar n\cdot (p_1-p_2)}{\bar n \cdot (p_1+p_2)},
\end{align}
where we have introduced a parameter $y\neq 1$, anticipating that (\ref{def}) is to be convoluted in $y$ with certain  distributions in a later section. 
In this setup the incoming and outgoing quark momenta are parametrized as 
\begin{align} \label{eq: nP}
\bar n\cdot p_1 = (y+\xi)\bar n\cdot P, \qquad \bar n\cdot p_2=(y-\xi)\bar n\cdot P. 
\end{align}
  As already suggested by (\ref{eq: nP}) and made clearer below, the physically relevant region is $\xi < y \le 1$. 
$W$ is the straight Wilson line along the $\bar{n}$ direction which makes the nonlocal operator gauge invariant\footnote{Our convention of the coupling constant is such that the covariant derivative reads $D^\mu = \partial^\mu -igA^\mu$.} 
\beq
 W(z_2,z_1) = P\exp\left(ig\int_{z_1}^{z_2}dz^\mu A_\mu(z)\right) =P\exp\left(-ig\int_{0}^{1}dv z^\mu_{12} A_\mu(z_{12}^v)\right) ,
\eeq
where
\begin{align}
z_{12}^\mu\equiv z_1^\mu -z_2^\mu,  \qquad (z_{12}^v)^\mu \equiv \bar{v} z_1^\mu  + vz_2^\mu, \qquad  \bar v \equiv 1-v.
\end{align}
Finally, the variable $t =\Delta^2$ with $\Delta^\mu= p_2^\mu-p^\mu_1$ constitutes the hard scale from our perspective. We assume $-t\gg -p^2_{1}\sim -p^2_2$ and introduce the power counting  parameter 
\begin{align}
\lambda \equiv \frac{\sqrt{-p^2}}{\sqrt{-t}} 
\ll 1 .
\label{lambda2}
\end{align} 
(We use the notation $p^2$ to represent  $p_1^2$ or $p_2^2$ when the distinction does not matter.) 

It should be mentioned that the quark-in-quark GPD (\ref{def}) is an unphysical object, since quark states are not proper asymptotic states of QCD. At this level, it is not entirely necessary to specify the scale of $p^2$, although naively $p^2\sim \Lambda^2_{\rm QCD}$ may seem a natural choice. In Sec.~\ref{proton new}, we will use (\ref{def}) as a building block of the   
physical (`quark-in-hadron') GPD at large-$t$, and at this point the scale $p^2$ needs to be carefully chosen. We expect that the hard dynamics governed by the scale $t$ decouples from the softer scale $p^2$ within the GPD,    
and the former is  fully captured in (\ref{def}). In other words, there should be a factorization formula
\begin{align}
F(x,y,\xi,t,p_1^2, p_2^2,\mu_{\rm UV}) = K(x,y,\xi,t,\mu_{\rm UV},\mu_F)  J(p_1^2,\mu_F) J(p_2^2,\mu_F)  S\left ( \frac{p_1^2 p_2^2}{-t}, \mu_F \right )    + {\cal O}(\lambda).  \label{f}
\end{align}
The  hard kernel  $K(x,y,\xi,t)$ is calculable in perturbation theory with $p^2_{1,2} = 0$ and subject to the usual subtraction of UV  divergences. It is understood that $K$ is a distribution in $y$ and resums the Sudakov double logarithms $(\alpha_s\ln^2(-t/\mu^2_F))^n$ along with subleading logarithms.  The IR divergences are  absorbed into the soft $S$ and jet $J$ functions to be specified later. As indicated by the notation, the left hand side does not depend on $\mu_F$, and the $\mu_{\rm UV}$-dependence is governed by the usual DGLAP/ERBL  equation for GPDs. The goal of this section  is to establish (\ref{f}) and study its RG properties. In the next section we will explicitly  calculate  $K(x,y,\xi,t)$ to one-loop.

\subsection{Breit frame}

In order to perform the power expansion in $\lambda$ (\ref{f}), we need to introduce a suitable light-cone basis that emphasize the role of $\sqrt{-t}$ as the hard scale. This will be different from the usual light-cone basis\footnote{We define the light-cone coordinates by  
\beq
x^\pm =x^0\pm x^3, \qquad x\cdot p= \frac{1}{2}(x^+p^-+x^-p^+) - \vec{x}_\perp\cdot \vec{p}_\perp. \label{lc}
\eeq
} 
in the GPD literature spanned by $\bar n^{\mu}= \delta^\mu_-\bar n^-$ (see (\ref{def})) and its conjugate $ n^{\mu}=\delta^\mu_+ n^+$.\footnote{
In the GPD literature, the vector $n^\mu$ typically points along the minus
direction. To comply with the standard notation in SCET, we have
instead defined $n^\mu$ to be along the plus direction
\cite{Schoenleber:2024ihr}.} 
Instead, we introduce light-like vectors $w^{\mu}, \bar w^{\mu}$, such that 
\beq
w^\mu \propto \lim_{p^2_1 \rightarrow 0} p^\mu_1, \qquad \bar w^\mu \propto \lim_{p_2^2 \rightarrow 0} p_2^\mu,  \label{breit}
\eeq
normalized as  
$w \cdot \bar w = 2$. 
The change of basis from $(n, \bar n)$ to $(w, \bar w)$ corresponds to a change of frames.  
The crucial difference is that in the $(n, \bar n)$ frame, $p_1$ and $p_2$ are roughly collinear in the same direction $\sim n$, while in the $(w, \bar w)$ frame, $p_1\sim w$ and $p_2\sim \bar w$ are collinear in opposite directions. The latter frame,  which is an analog of the Breit frame, can be achieved from the former via a series of Lorentz transformations as described in Appendix~\ref{app A}. 
 It should be emphasized that the quark-in-quark GPD (\ref{def}) is invariant under Lorentz transformations (provided  $\bar{n}^\mu$ is transformed in the same way as $p^\mu_1,p_2^\mu$) and can therefore be analyzed in any convenient frame. In the next subsections, we will establish the factorization of $F$ in the $(w,\bar w)$ frame. The result is then valid in the original  $(n,\bar n)$ frame.

In the $(w,\bar w)$ frame,  it is possible to take
\begin{align}
\bar n^{\mu} = \bar w^{\mu} + w^{\mu} + \bar n_{\perp}^{\mu}, \qquad \bar n_\perp^2=-4,  \label{nnn}
\end{align}
 by exploiting the fact that (\ref{def}) is invariant under rescaling $\bar n^\mu \to c \bar n^\mu$, see Appendix~\ref{app A}. 
 The quark momenta can now be written as
\begin{align}
p_1^{\mu} = \bar w \cdot p_1 \frac{w^{\mu}}{2} + \frac{p_1^2}{\bar w \cdot p_1} \frac{\bar w^{\mu}}{2},
\qquad 
p_2^{\mu} = w \cdot p_2 \frac{\bar w^{\mu}}{2} + \frac{p_2^2}{w \cdot p_2} \frac{w^{\mu}}{2},
\end{align}
where $\bar w \cdot p_1 \sim w \cdot p_2 
\sim \sqrt{-t}$. We can then write 
\begin{align} \label{wp}
\bar w \cdot p_1 &= \frac{1}{2} (y+ \xi) (\bar n \cdot P) + \frac{1}{2} \sqrt{(y+\xi)^2 (\bar n \cdot P)^2 - 4 p^2} = (y+ \xi) (\bar n \cdot P) + \f O(\lambda^2), 
\\ \label{wp_2}
w \cdot p_2 &= \frac{1}{2} (y- \xi) (\bar n \cdot P) + \frac{1}{2} \sqrt{(y-\xi)^2 (\bar n \cdot P)^2 - 4 p_2^2} = (y- \xi) (\bar n \cdot P) + \f O(\lambda^2).
\end{align}
Since we are only interested in logarithms of $p^2_{1,2}$  that arise from IR regularization, we may set $p^2_{1,2}  = 0$ whenever it is safe to do so. 
Using 
\begin{align}
- t =  \bar w \cdot p_1 \, w \cdot p_2 + \f O(p^2_{1,2}) = (y^2 - \xi^2) (\bar n \cdot P)^2 + \f O(\lambda^2),
\end{align}
we find
\begin{align}
\bar n \cdot P = \sqrt{\frac{-t}{y^2 - \xi^2}} + \f O(\lambda^2)
\end{align}
and therefore, 
\begin{align}
p_1^{\mu} = \sqrt{\frac{y+\xi}{y-\xi} } \sqrt{-t}  \frac{w^{\mu}}{2} + \f O(\lambda^2), \qquad 
p_2^{\mu} = \sqrt{\frac{y-\xi}{y+\xi} } \sqrt{-t}  \frac{\bar w^{\mu}}{2} + \f O(\lambda^2). \label{xiy}
\end{align}
Note that, since we are interested in large negative values of $t$, we must require that $\xi<y\le 1$.\footnote{The region $-1<y<-\xi$ is associated with the anti-quark GPD and will not be considered in this paper. We are only interested in the region $y\approx 1$, as we explain later. }  
In fact, the change of frames $(n,\bar n)\to (w,\bar w)$ in Appendix A is meaningful only in this range.

\subsection{Soft Collinear Effective Theory}

Our main theoretical tool to analyze the quark-in-quark GPD (\ref{def}) is  the Soft Collinear Effective Theory (SCET) in its position space formulation \cite{Beneke:2002ph, Beneke:2002ni}. As in the case of the quark electromagnetic form factor \cite{Becher:2014oda}, we may use $\rm SCET_I$.  The setup of the effective theory is standard, so we only recall  the relevant definitions. 
Let us decompose a  generic vector $k^{\mu}$ in the basis $(w,\bar{w})$ as
\beq
k^\mu =\frac{\bar{w}\cdot k}{2} w^\mu + \frac{w\cdot k}{2} \bar{w}^\mu + k^\mu_\perp \equiv (\bar{w}\cdot k, w\cdot k,k_\perp),
\eeq 
and introduce the collinear $(c)$, anti-collinear $(\bar{c})$ and ultrasoft ($us$)  modes
\begin{align}
k^\mu_c \sim (1, \lambda^2, \lambda) \sqrt{-t}, \qquad k^\mu_{\bar c} \sim (\lambda^2, 1, \lambda) \sqrt{-t}, \qquad k_{us}^\mu \sim (\lambda^2, \lambda^2, \lambda^2) \sqrt{-t}, \label{region}
\end{align}
respectively.\footnote{The soft mode $k^\mu_s\sim (\lambda,\lambda,\lambda)\sqrt{-t}$ does not explicitly appear in the massless Sudakov problem within SCET \cite{Becher:2014oda}.}  
In accordance with (\ref{region}),  we decompose  the quark and gluon fields as 
\begin{align} \label{eq: mode separation}
\psi = \psi_c + \psi_{\bar c} + \psi_{us}, \qquad A^\mu = A_c^\mu + A^\mu_{\bar c} + A^\mu_{us}. 
\end{align}
The collinear and anti-collinear quark fields 
 are further projected onto the two leading spinor components
\begin{align} \label{eq: spinor}
\xi_c = \frac{\s w \sbar w}{4} \psi_c, \qquad \xi_{\bar c} = \frac{\sbar w \s w}{4} \psi_{\bar c}.
\end{align}
In Fourier space, the collinear fields $\xi_c,A_c$ carry momentum of order $k_c$, and similarly for the anti-collinear $\xi_{\bar{c}},A_{\bar{c}}\sim k_{\bar{c}}$ and ultrasoft fields  $\psi_{us},A_{us}\sim k_{us}$. This implies specific counting rules for different components of the  fields 
\beq
\xi_c,\xi_{\bar{c}} \sim \lambda, \qquad \psi_{us}\sim \lambda^3, \qquad  
A_c^\mu \sim k_c^\mu, \qquad A_{\bar{c}}^\mu \sim k_{\bar{c}}^\mu, \qquad  A_{us}^\mu\sim k_{us}^\mu.
\eeq
Derivatives $\partial^\mu \sim k^\mu$ acting on fields are typically suppressed by $\lambda$ except for projections along large momentum  components.
The ultrasoft quark fields $\psi_{us}$ are irrelevant  at leading power, and the ultrasoft gluon fields $A_{us}$ can be factored out by the so-called decoupling transformation
\begin{align} \label{eq: decoupling}
\xi_c(x) \rightarrow Y_w\left ( \frac{\bar w \cdot x}{2} w \right ) \xi_c(x), \qquad A^\mu_c(x) \rightarrow Y_w\left ( \frac{\bar w \cdot x}{2} w \right ) A^\mu_c(x) Y_w^{\dagger} \left ( \frac{\bar w \cdot x}{2} w \right ),
\end{align}
and similarly with $(c,w) \leftrightarrow (\bar c, \bar w)$, where the ultrasoft Wilson lines are defined by
\begin{align}
Y_w(x) = P \exp \left ( ig \int_{-\infty}^0 ds\, w \cdot A_{us}(x + s w) \right ).
\end{align}
After this transformation, gauge invariance can be imposed  separately in the collinear, anticollinear and ultrasoft sectors. 
The leading twist quark operators are comprised of gauge invariant combinations in the collinear and anti-collinear sectors  
\begin{align}
\chi_c(x) \equiv W_c^{\dagger}(x) \xi_c(x), \qquad \chi_{\bar c}(x) \equiv W_{\bar c}^{\dagger}(x) \xi_{\bar c}(x),
\end{align}
where 
\beq
W_c(x) \equiv\! P \exp \left ( ig \int_{-\infty}^0 
\!ds\, \bar w \cdot A_c(x + s \bar w) \right ), \ \   
W_{\bar c}(x) \equiv\! P \exp \left ( ig \int_{-\infty}^0 \!ds\, w \cdot A_{\bar c}(x + s w) \right )\!.
\eeq

\subsection{Factorization}

Let us first recall that the standard factorization of the Sudakov form factor follows from the SCET operator expansion
\begin{align} \notag
\bar \psi(x) \gamma^{\mu} \psi(x) &= \int^\infty_{-\infty} ds_1 ds_2\, \w C(s_1,s_2,\mu_F) \bar \chi_{\bar c}\left (\bar w \cdot x \, \frac{w}{2}+s_2 w \right ) \gamma_{\perp}^{\mu} Y_{\bar w}^{\dagger}\left ( 0\right ) Y_{w}\left ( 0 \right ) \chi_{c}\left ( w \cdot x \, \frac{\bar w}{2} +s_1 \bar w \right )
\\  \label{eq: SCET op exp} 
&\quad + (c,w \leftrightarrow \bar c, \bar w) + \f O(\lambda^3),
\end{align} 
where it is understood that matrix elements with respect to the incoming collinear and outgoing anti-collinear quark states are taken.
The local operator is mapped to a non-local operator because the  derivative operators  $(\bar w\cdot \partial)^n \chi_c$, $(w\cdot \partial)^n \bar{\chi}_{\bar c}$ are not power-suppressed. The hard 
 `matching' kernel  $\w C$, or its Fourier transform $C$ has been calculated  in perturbation theory up to four loops \cite{Lee:2022nhh}. To one-loop, the result is well known 
\begin{align} \notag
C(t,\mu_F) &= \int ds_1ds_2\, \w C(s_1,s_2) e^{is_2 w \cdot p_2 - i s_1 \bar w \cdot p} 
\\  \label{eq: local hard}
&=  1+ \frac{\alpha_s C_F}{4\pi} \left ( - \ln^2 \frac{\mu^2_F}{-t} - 3 \ln \frac{\mu^2_F}{-t} - 8 + \frac{\pi^2}{6} \right ) + \f O(\alpha_s^2). 
\end{align}
After using the factorization of Hilbert spaces at leading power into collinear, anti-collinear and ultrasoft sectors, one obtains
\begin{align}
\bra{p_2} \bar \psi(x) \gamma^{\mu} \psi(x) |p_1\rangle = C(t,\mu_F) \bra{p_2} \bar \chi_{\bar c}(x) \ket 0 \gamma_{\perp}^{\mu} \bra 0 Y_{\bar w}^{\dagger}\left ( 0\right ) Y_{w}\left ( 0 \right ) \ket 0 \bra 0 \chi_{c}(x) |p_1\rangle + \f O(\lambda), \label{factori}
\end{align} 
where we have translated the quark fields and used the fact that $|p_{1,2}\rangle \sim \lambda^{-1}$. For the nucleon states, the situation is more complicated, since hadronic matrix elements do not admit such a simple factorization between ultrasoft and collinear modes.  Furthermore, the power counting is not anymore manifest, so that higher power operators neglected in  \eqref{eq: SCET op exp} might be  more important than the leading operator $\bar \chi_{\bar c} Y^\dagger_{\bar w} Y_w \chi_c$. 
We will discuss this issue in later sections.

For the moment, let us continue to discuss the case of external quark states. 
In the non-local case \eqref{before}, we will argue that (\ref{eq: SCET op exp}) is generalized as
\begin{align} \label{eq: factorization 1}
& \bar \psi(z_2\bar n) \sbar n W(z_2\bar n,z_1\bar n) \psi(z_1 \bar n) \nn
&= \int ds_1 ds_2 \, \w K(s_1 , s_2,z_1-z_2) \, \bar \chi_{\bar c}((z_2+s_2) w) \sbar n_{\perp} Y_{\bar w}^{\dagger}\left ( 0\right ) Y_{w}\left ( 0 \right ) \chi_{c}((z_1+s_1) \bar w) 
\nn & \qquad 
+  (c,w \leftrightarrow \bar c, \bar w)+\f O(\lambda^3). 
\end{align}
This follows from the general effective field theory principle that the complete operator basis is constructed by considering all possible operators that have the appropriate symmetry properties. Since each additional operator building block introduces a power suppression, the minimum (and hence leading) operator content that has a non-zero overlap with the quark states is $\bar \xi_{\bar c}(z_2\bar{n}) \xi_c(z_1\bar{n}) = \bar \xi_{\bar c}(z_2w) \xi_c(z_1\bar w) + \f O(\lambda^3)$, which must then be dressed with the appropriate collinear, anticollinear and ultrasoft Wilson lines to ensure gauge invariance. In other words, the minimal combination of operator building blocks $\bar \chi_{\bar c} \chi_c$ is the same regardless of whether the hard current is local or non-local.

This simple argument is rather abstract and one might be concerned about possible issues related to the factorization of the Wilson line $W(z_2\bar n,z_1\bar n)$. To further substantiate \eqref{eq: factorization 1}, we provide another more technical argument in the following. 
First, we see from  \eqref{eq: mode separation} that the initial Wilson line in the $\bar n$ direction involves the component 
\begin{align} \label{eq: barn A approx}
\bar n \cdot A( z \bar n) = \bar w \cdot A_c ( z \bar w) + w \cdot A_{\bar c}(z w) + \f O(\lambda),
\end{align}
where we  neglected the ultrasoft field as well as the dependence on transverse coordinates since they are power-suppressed.
We can then approximate the Wilson line as 
\begin{align} \notag
W(z_2\bar n,z_1\bar n) &= W(z_2(w+\bar w),z_1(w+\bar w)) + \f O(\lambda) 
\\ \label{eq: Wline relation 1}
&= W_{\bar c}(z_2w,z_1w)W_c(z_2\bar w,z_1\bar w) + \f O(\lambda).
\end{align}
The first step follows from (\ref{eq: barn A approx}). 
The second step follows from the non-Abelian Stokes theorem \cite{Halpern:1978ik,Arefeva:1979dp} applied to the triangular contour formed by the vertices $z_2(w+\bar{w})$, $z_1w+z_2\bar{w}$ and $z_1(w+\bar{w})$ 
\beq \label{eq: Wline relation 2}
W(z_2(w+\bar{w}),z_1(w+\bar{w}))&=&W(z_2(w+\bar{w}),z_1w+z_2\bar{w})W(z_1w+z_2\bar{w},z_1(w+\bar{w}))  \nn && \times P\exp\left(\frac{i}{2}\int_{\cal A} \frac{d(w\cdot x) d(\bar{w}\cdot x)}{4} UF^{\mu\nu}(x)U^\dagger w_\mu \bar{w}_\nu\right) \nn
&=&  W_{\bar c}(z_2 w,z_1 w)W_c(z_2 \bar w,z_1 \bar w) + \f O(\lambda^2). 
\eeq
In the exponential factor, the integral is over the area ${\cal A}$ of the triangle and $U$ is a Wilson line connecting an arbitrary point on the contour to the interior point $x$. (The result does not depend on the choice of $U$.) Since  $F^{\mu\nu}w_\mu \bar{w}_\nu \sim \lambda^2$, this factor is $1 + \f O(\lambda^2)$.  The remaining Wilson lines are then along the $w$ and $\bar w$ directions respectively, so that we can approximate the corresponding gluon fields in the path ordered exponential by $A_{\bar c}$ and $A_c$, expanded around their respective light-cones.
The resulting expression must be made gauge invariant by attaching the collinear/anti-collinear Wilson line
\begin{align} \notag
W_c(z_2 \bar w,z_1 \bar w)\xi_c(z_1\bar{w}) &\to W_c(-\infty \bar w,z_2\bar{w}) 
W_c(z_2 \bar w,z_1 \bar w) \xi_c(z_1\bar{w}) 
\\ \notag
&\qquad = W_c^\dagger(z_1\bar w)\xi_c(z_1\bar w) = \chi_c(z_1\bar w),
\\ \label{eq: repl 1}
\bar \xi_{\bar c}(z_2w)W_{\bar c}(z_2w,z_1w) &\to \bar \xi_{\bar c}(z_2w)W_{\bar c}(z_2w,z_1w) W_{\bar c}(z_1w,-\infty w) 
\\ \notag
&\qquad =\bar{\xi}_{\bar c}(z_2w)W_{\bar c}(z_2w)=\bar{\chi}_{\bar c}(z_2w).
\end{align}
Graphically, the Wilson line $W_c(-\infty \bar w,z_2\bar{w})$ ($W_{\bar c}(z_1w,-\infty w)$) originates from  the resummation and factorization of 
collinear gluons attaching to the anti-collinear quark leg (anti-collinear gluons attaching to the collinear quark leg). Of course, the operator relations \eqref{eq: Wline relation 1} and \eqref{eq: Wline relation 2} are only formal in the sense that we have not taken into account the effect of integrating out the hard modes. This  should result in the 
hard factor $\widetilde{K}$ convoluted along the collinear and anti-collinear directions as in \eqref{eq: SCET op exp}. In the nonlocal case, it will  additionally depend on the difference $z_1-z_2$ so that \eqref{eq: SCET op exp} is recovered in the local limit. 
Finally, we must  apply the decoupling transformation \eqref{eq: decoupling}, producing  the ultrasoft Wilson lines $Y_w,Y^\dagger_{\bar{w}}$. We thus arrive at  \eqref{eq: factorization 1}.

Since the collinear and anti-collinear sectors of the theory are decoupled at leading power (after the decoupling transformation \eqref{eq: decoupling}), the quark matrix element of (\ref{eq: factorization 1}) factorizes 
\begin{align}
\bra {p_2} \bar \chi_{\bar c} Y_{\bar w}^{\dagger} Y_w \chi_c \ket{p_1} = \bra{p_2} \bar \chi_{\bar c} \ket 0 \, \bra{0} Y_{\bar w}^{\dagger} Y_w \ket 0\, \bra 0 \chi_c \ket{p_1} + \f O(\lambda).
\end{align}
Performing the Fourier transform, we obtain \eqref{f} with the identifications 
\beq
&&K(x,y,\xi,t, \mu_F) = \frac{ \bar n \cdot P }{2\pi}  \int_{-\infty}^{\infty} ds_1 d s_2 dz\,  e^{i(x-y)  z\bar n \cdot P + is_2 w \cdot p_2 -is_1 \bar w \cdot p_1} \w K(s_1 , s_2 , z, \mu_F), \label{kdef}
\\
&& J(p^2_1, \mu_F) u(p_1) = \bra 0 \chi_c(0) \ket{p_1}, \qquad \bar{u}(p_2)J(p_2^2,\mu_F) = \langle p_2|\bar{\chi}_{\bar{c}}|0\rangle, \label{jdef}
\\
&& S\left ( \frac{p^2_1 p_2^2}{-t} , \mu_F \right ) = \bra 0 Y_w^{\dagger}\left ( 0\right ) Y_{\bar w}\left ( 0 \right ) \ket 0.  \label{sdef}
\eeq
Note that the operator definition of $S$ is actually scaleless but in our case it depends on $\frac{p^2_1 p_2^2}{-t\mu_F^2}$ from the off-shell regularization  $p_{1,2}^2\neq 0$ of the IR divergences. This is a well-known complication that arises also in the local case (\ref{eq: SCET op exp}).

\subsection{RG equations}
\label{rgeq}
We shall now resum the Sudakov logarithms $\ln^2 \frac{-t}{\mu_F^2}$ by using RG equations. 
The factorization scale $\mu_F$ is an artificial parameter that separates physics between the scales $-t$ and $p^2_{1,2}$. The total GPD $F = KJJS$ does not depend on this scale, but one can resum large logarithms separately in $K$ and $JJS$. In fact, since the combination $JJS$ is  the same  as in the local case, $K$ has the same evolution equation in $\mu_F$ as $C$ for the local operator \eqref{eq: local hard}, and the  anomalous dimensions of the latter are  known to  four-loops \cite{vonManteuffel:2020vjv}. In essence, the non-locality merely concerns the hard dynamics at the scale $-t$ and the IR physics is not influenced by the non-locality.

Performing the resummation is a standard  material, but we repeat the main expressions here. 
Suppose now we evaluate \eqref{f} at $\mu_F\sim \sqrt{-t}$. This choice minimizes the large logarithms $\ln^2 \frac{-t}{\mu_F^2}$ in the hard function $K$. 
We can then evolve down the factors $JJS$ from $\mu_F\sim \sqrt{-t}$ to $\mu_0\sim  \sqrt{-p^2}$ in order to resum the logarithms\footnote{ Note that the soft function $S$ still contains large logarithms of $\frac{(p^2)^2}{t\mu_0^2}\sim \frac{p^2}{t}$. They can be independently resummed by the renormalization group evolution of $S$. For our discussion in later sections, only the factor $U$ is needed. } 
\beq
J(p^2_1, \mu_F) J(p_2^2, \mu_F) S\left ( \frac{p^2_1 p_2^2}{-t}, \mu_F \right ) = U(\mu_F,\mu_0) J(p^2_1, \mu_0) J(p_2^2, \mu_0) S\left ( \frac{p_1^2 p_2^2}{-t}, \mu_0 \right ), \label{rg}
\eeq
where
\begin{align}
U(\mu_F, \mu_0) = \exp \left ( 2{\cal S}(\mu_F,\mu_0) - a_{\gamma_C}(\mu_F,\mu_0) \right ) \left ( \frac{-t}{\mu_F^2}\right )^{- a_{\Gamma_{\rm cusp}}(\mu_F , \mu_0)}, \label{cu}
\end{align}
is the well-known Sudakov exponential. 
We have
\begin{align}
{\cal S}(\mu_F, \mu_0) &= - \int_{\alpha_s(\mu_F)}^{\alpha_s(\mu_0)} d\alpha \, \frac{\Gamma_{\rm cusp}(\alpha)}{\beta(\alpha)} \int_{\alpha_s(\mu_F)}^{\alpha} \frac{d\alpha'}{\beta(\alpha')}
\\
a_{\gamma_C}(\mu_F,\mu_0) &= - \int_{\alpha_s(\mu_F)}^{\alpha_s(\mu_0)} d\alpha \, \frac{\gamma_C(\alpha)}{\beta(\alpha)},
\end{align}
where $\beta(\alpha) = \frac{\alpha^2}{\pi} \left ( - \frac{11}{6} C_A + \frac{2}{3} n_f T_F  \right ) + \f O(\alpha^3)$ is the QCD beta function, $\Gamma_{\rm cusp}(\alpha) = \frac{\alpha C_F}{\pi} + \f O(\alpha^2)$ is the cusp anomalous dimension and $\gamma_C(\alpha) =  \frac{\alpha C_F}{\pi} ( - \frac{3}{2} ) + \f O(\alpha^2)$ is the non-logarithmic part of the anomalous dimension of the local hard function $C$ in \eqref{eq: local hard}. 
Equivalently, since $KJJS$ is $\mu_F$-independent,  (\ref{rg}) can be viewed as the evolution of $K$ 
\beq
K(t,\mu_{\rm UV},\mu_0)=K(t,\mu_{\rm UV},\mu_F)U(\mu_F,\mu_0), \label{devolve}
\eeq
which dictates the $t$-dependence of $K$ after the Sudakov resummation.

\section{One-loop calculation using the method of regions} \label{sec: 1loop calc}

In this section, we corroborate the factorization formula (\ref{f}) in the previous section by explicit calculation.
We evaluate the quark GPD $F(x,y, \xi,t)$ (\ref{def}) to one-loop in perturbation theory and determine all the factors $K,J,J,S$ to this order. For this purpose, we employ the  method of regions \cite{Beneke:1997zp} which is a standard tool in SCET and multi-loop calculations. This method has also been proven highly effective in calculating the electromagnetic form factor $F(t)=\int dx F(x,\xi,t)$, see, e.g.,  \cite{Smirnov:2002pj,Becher:2014oda,Anastasiou:2020vkr}. Here we generalize this analysis to the quark-in-quark GPD.

\begin{figure}[t]
\vspace{-5mm}
    \centering
\includegraphics[width=1.1\linewidth]{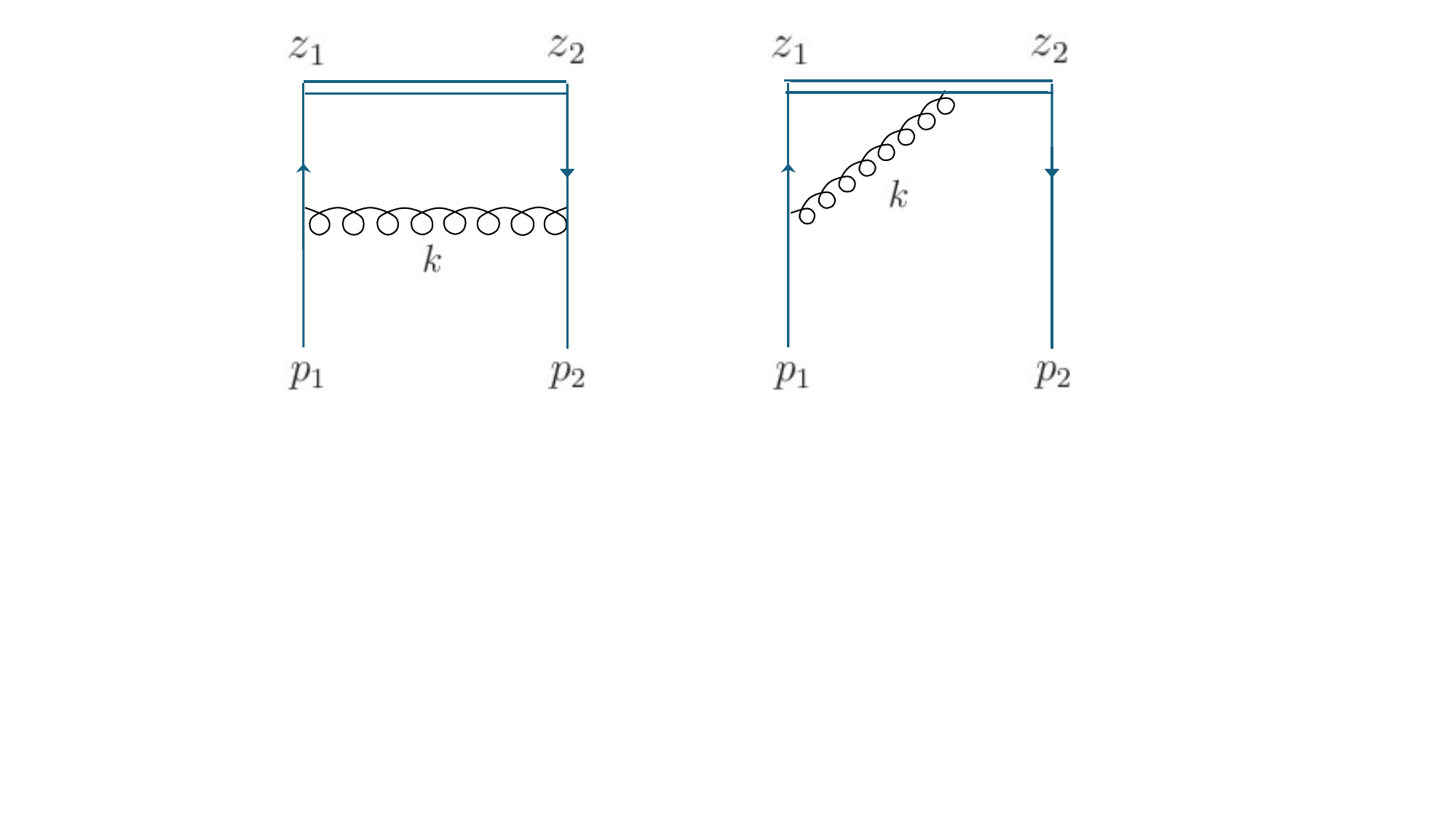}
\vspace{-55mm}
    \caption{Left: Box diagram. Right: Sail diagram  }
    \label{diagrams}
\end{figure}

At the tree level,
\beq
F(x,y,\xi,t) = \begin{cases}  \delta(x-y), \qquad (x>\xi) \\
0. \qquad \qquad (x<\xi) \end{cases}
\eeq
Note that $F$ vanishes in the ERBL region $x<\xi$ to this order due to  our assumption $y>\xi$ (see the comment after (\ref{xiy})). 
To one-loop in $d=4-2\epsilon$ dimensions, and before the Fourier transform (see (\ref{before})), 
the integrals to be evaluated  are 
the `box' diagram (see Fig.~\ref{diagrams})
\beq
\tilde{F}^{\rm box}(z_1,z_2,p_1,p_2) = -i g^2C_F\int \frac{d^dk}{(2\pi)^d} \frac{\bar u(p_2) \gamma^{\mu} (\Slash k + \Slash p_2) \Slash n (\Slash k + \Slash p_1) \gamma_{\mu} u(p_1)}{k^2 (k+p_1)^2 (k+p_2)^2} e^{iz_2\cdot (k+p_2 ) - iz_1\cdot (k+p_1)}, \label{box}
\eeq
 the `sail' diagram 
\begin{align}
 \tilde{F}^{\rm sail} (z_1,z_2,p_1,p_2)=&  g^2C_F\int_0^1 dv \int \frac{d^dk}{ (2\pi)^{d}} e^{iz_2 \cdot p_2 -iz_1 \cdot(k+p_1) + iz_{12}^v \cdot k}\frac{\bar u(p_2) \Slash n (\Slash p_1 + \Slash k) \Slash z_{12} u(p_1)}{k^2 (p_1+k)^2} \nn  &+({\rm mirror\ diagram}), \label{sail}
\end{align}
and the self-energy diagram 
\beq
\tilde{F}^{\rm self}(z_1,z_2,p_1,p_2)&=&-i\frac{g^2C_F}{2} e^{iz_2 \cdot p_2 -iz_1 \cdot p_1 }  \!\int \!\frac{d^dk}{ (2\pi)^{d}}\frac{ \bar u(p_2) \Slash n \Slash p_1\gamma_\mu (\Slash p_1+\Slash k)\gamma^\mu u(p_1) }{p_1^2 k^2 (p_1+k)^2} \nn && +({\rm mirror\ diagram}) ,  \label{self}
\eeq
where the factor of $\frac{1}{2}$ is from the LSZ reduction formula. 
In the method of regions, one   divides the loop momentum integral   into four regions 
\beq
\int d^dk = \int_h d^dk + \int_c d^dk + \int_{\bar c} d^dk + \int_{us} d^dk, \label{4}
\eeq
where the hard $(h)$ region is characterized by hard modes  $k^\mu_h\sim (1,1,1)\sqrt{-t}$ and the other regions are as in (\ref{region}). 
In accordance with (\ref{4}), we write these integrals as 
\beq
F^a=F^a_h+F^a_c+F^a_{\bar c}+F^a_{us}. \quad 
\eeq
In each region, different approximations are to be used. In the hard ($h$) region, we set  $p^2_1=p_2^2=0$. 
In the collinear ($c$) region, we may approximate 
\beq
(k+p_2)^2 \approx 2k\cdot p_2 \approx \bar{w}\cdot k w\cdot p_2, \quad n\cdot (k+p_2) \approx \bar{w}\cdot k + w\cdot p_2,\quad   n\cdot (k+p_1) \approx \bar{w}\cdot (k+p_1),
\eeq
while $(k+p_1)=k^2+2k\cdot p_1+p_1^2$ is treated in full. 
Similarly, in the anti-collinear ($\bar{c}$) region 
\beq
(k+p_1)^2 \approx 2k\cdot p_1 \approx w\cdot k \bar{w}\cdot p_1, \quad n\cdot (k+p_2) \approx w\cdot (k+ p_2),\quad   n\cdot (k+p_1) \approx w\cdot k+\bar{w}\cdot p_1,
\eeq
and no approximation for  $(k+p_2)^2=k^2+2k\cdot p_2+p_2^2$. 
In the ultrasoft ($us$) region, we may set $k=0$ in the numerator and 
\beq
(k+p_1)^2\approx p^2_1+w\cdot k \bar{w}\cdot p_1, \qquad (k+p_2)^2 \approx p_2^2 + \bar{w}\cdot k  w\cdot p_2,
\eeq
in the denominator.

\subsection{Hard region}

The hard contribution $F_h$ with $t\neq 0$ has been computed in \cite{Bhattacharya:2023wvy} setting $y=1$. Our calculation, besides generalizing to the case  $y\neq 1$, is slightly different in the following sense. In \cite{Bhattacharya:2023wvy}, the quark GPD $F(x,y=1)$ was treated as a distribution in $x$ to be convoluted with the hard kernel in Compton scattering. For a reason to become clear in later sections, here we treat $F(x,y)$ as a distribution in $y$. We comment on the connection to the result \cite{Bhattacharya:2023wvy} at  the end.   For notational simplicity, in the following we set  the UV renormalization and factorization scales to be equal $\mu_{\rm UV}=\mu_F\equiv \mu$ in intermediate expressions. The distinction will be restored in the final results. 

Let us start with the box diagram (\ref{box}). After introducing Feynman parameters and shifting $k^\mu$, we can perform the Fourier transform  
\begin{align}
\bar{n}\cdot P \int_{-\infty}^{\infty} \frac{dz}{2\pi} e^{izx \bar{n}\cdot P}  e^{-iz_{12}^u p_1 + iz_{21}^v p_2} = \delta \left ( x - (1-  u - v)y + (u-v) \xi\right ). 
\end{align}
This constrains the integration over the Feynman parameters $u,v$ as 
\begin{align} 
&\int_0^1 du \int_0^{\bar u} dv \, \delta \left ( x - (1- u - v) y +  (u-v) \xi\right )
= \frac{\theta(x-|\xi|) \theta(y-x) }{y-|\xi|} \int_0^{\frac{y-x}{|\xi| + y}}du \\   &\qquad +\frac{\theta(-x-|\xi|)\theta(x-y) }{|\xi| - y} \int_0^{\frac{x-y}{-|\xi| - y}} du + \frac{ \theta(|\xi| - |x| ) }{|\xi| - y} \Bigg \{  \theta(y-x) \int_{\frac{y-x}{|\xi| + y}}^{\frac{|\xi| - x}{2|\xi|}} du \, + \theta(x-y) \int_0^{\frac{|\xi| - x}{2|\xi|}} du \,  \Bigg \}, \notag
\end{align}
valid  $\forall x,y,\xi \in \mathbb R$. As usual, we restrict ourselves to the case $\xi > 0$.
We can thus write    
\begin{align}
F^{\rm box}_h(x,y,\xi,t) = \frac{\alpha_s C_F}{2\pi} 
\theta(x+\xi)\left(\frac{\mu^2}{-t}\right)^\epsilon \Bigl( \theta(x-\xi)  F^{\rm box}_{h,\rm DGLAP}  + \theta(\xi - x)  F^{\rm box}_{h,\rm ERBL} \Bigr).
\end{align}
The remaining integrals are straightforward. In the DGLAP region $x>\xi$, we find  
\begin{align}
&F^{\rm box}_{h,\rm DGLAP}= -\frac{\delta(y-x)}{\epsilon^2_{\rm IR}} +\frac{1}{\epsilon_{\rm UV}} \frac{y-x}{y^2-\xi^2} + \frac{1}{\epsilon_{\rm IR}}\left(\frac{2(xy-\xi^2)}{y^2-\xi^2}\left[\frac{\theta(y-x)}{y-x}\right]_+ +\delta(x-y)\ln \frac{(1-x)^2}{y^2-\xi^2}\right) \nn 
&\quad +\frac{(x^2+y^2-2\xi^2)\ln (y^2-\xi^2)-(x-y)^2}{y^2-\xi^2}\left[\frac{\theta(y-x)}{y-x}\right]_+ -2\frac{x^2+y^2-2\xi^2}{y^2-\xi^2}\left[\frac{\theta(y-x)\ln(y-x)}{y-x}\right]_+ \nn 
&\quad  +\delta(y-x) \left(-\frac{1}{2}\ln^2\frac{y^2-\xi^2}{(1-x)^2}+\frac{\pi^2}{12}\right) .\label{dg}
\end{align}
The terms proportional to the delta function $\delta(x-y)$ are different from the corresponding expression in \cite{Bhattacharya:2023wvy} (even after setting $y=1$) because the meaning of the plus prescription is different. To obtain (\ref{dg}), we have used the formula\footnote{This follows from the identity 
\beq
\int_{-1}^1 dy \frac{\theta(y-x)}{(y-x)^{1+2\epsilon}}g(y)&=& -\frac{(1-x)^{-2\epsilon}}{2\epsilon} g(x) + \int_x^1 \frac{dy}{(y-x)^{1+2\epsilon}}(g(y)-g(x)) \nn 
&=& -\frac{1}{2\epsilon}g(x) + \ln (1-x)g(x) + \int_x^1 \frac{dy}{y-x}(g(y)-g(x))+{\cal O}(\epsilon). \label{testf}
\eeq
}
\beq
\frac{\theta(y-x)}{(y-x)^{1+2\epsilon}} &=& -\frac{\delta(y-x)}{2\epsilon} +\left[\frac{\theta(y-x)}{y-x}\right]_+ + \ln (1-x)\delta(y-x) \nn &&+ \epsilon\left( -\ln^2(1-x)\delta(y-x) -2\left[\frac{\theta(y-x)\ln(y-x)}{y-x}\right]_+\right)+{\cal O}(\epsilon^2).
\eeq
It is understood that (\ref{dg}) is to be convoluted with a test function $g(y)$ with the usual rule  
\beq
\int dy \left[\frac{\theta(y-x)}{y-x}\right]_+ g(y) = \int_x^1 dy\frac{g(y)-g(x)}{y-x}.
\eeq

In the ERBL region $-\xi<x<\xi$, in principle $F_h$ has support in both regions $x<y$ and $x>y$. However, we have already argued that the only physically relevant region is $x<\xi<y$. We thus restrict to this case and obtain 
\begin{align}
F^{\rm box}_{h,\rm ERBL}=& \frac{x+\xi}{y+\xi}\left(\frac{1}{2\xi}\frac{1}{\epsilon_{\rm UV}}+\frac{1}{y-x}\frac{1}{\epsilon_{\rm IR}}\right)  -\frac{x+\xi}{2\xi(y+\xi)} \\
& + \frac{1}{2\xi (y-x)(y^2-\xi^2)}\Biggl[ (x-y)(x+\xi)(y+\xi)\ln \frac{\xi^2-x^2}{4\xi^2}  +2\xi x^2 \ln \frac{2\xi(y+\xi)}{(x+\xi)(y-x)} \nn  &  + 2\xi^3\ln \frac{\xi+x}{\xi-x}+2\xi y^2 \ln \frac{(y+\xi)(\xi-x)}{2\xi(y-x)}+4\xi^3\ln \frac{y-x}{y+\xi}\Biggr] . \nonumber
\end{align}
This agrees with \cite{Bhattacharya:2023wvy} after setting $y=1$. 

Next consider the hard region of the sail diagram (\ref{sail}) which  reduces to a scaleless integral\footnote{In the SCET literature, scaleless integrals are often immediately  discarded by setting $\epsilon_{\rm UV}=\epsilon_{\rm IR}$. However, in our one-loop calculation we can keep the distinction with little extra effort. This allows us to explicitly check the cancellation of $\frac{1}{\epsilon_{\rm IR}}$ poles. }  
\beq
F_{h}^{\rm sail}=\frac{\alpha_s C_F}{2\pi}
\left(\frac{1}{\epsilon_{\rm UV}}-\frac{1}{\epsilon_{\rm IR}}\right)  (f(x,y,\xi)+f(x,y,-\xi)),
\eeq
where 
\beq
f(x,y,\xi)&=&\int_0^1 du\frac{1-u}{u}\left(\delta(x-y+u(y+\xi)) -\delta(x-y)\right) \nn 
&=&\int_0^1 du\frac{1}{u}\left(\delta(x-y+u(y+\xi)) -\delta(x-y)\right) -\frac{\theta(y-x)}{y+\xi}+\delta(x-y) 
\eeq
and $f(x,y,-\xi)$ is from the mirror diagram. 
Using 
\beq
\lim_{\delta\to 0} \int_0^1 \frac{du}{u^{1+\delta}}\left(\delta(x-y+u(y+\xi)) -\delta(x-y)\right) =\left[\frac{\theta(y-x)}{y-x}\right]_+ +\delta(x-y)  \ln \frac{1-x}{x+\xi}
\eeq
we find 
\beq
F_{h,{\rm DGLAP}}^{\rm sail}&=&\frac{\alpha_s C_F}{2\pi}
\left(\frac{1}{\epsilon_{\rm UV}}-\frac{1}{\epsilon_{\rm IR}}\right) \nn && \times \left(\frac{2(xy-\xi^2)}{y^2-\xi^2}\left[\frac{\theta(y-x)}{y-x}\right]_+ + \delta(x-y)\left(2+ \ln \frac{(1-x)^2}{y^2-\xi^2} \right)\right). \label{sailh}
\eeq
 In the ERBL region $x<\xi<y$, the second term $f(x,y,-\xi)$ vanishes because $u=\frac{y-x}{y-\xi}>1$. We get 
\beq
F^{\rm sail}_{h,{\rm ERBL}} = \frac{\alpha_s C_F}{2\pi}
\left(\frac{1}{\epsilon_{\rm UV}}-\frac{1}{\epsilon_{\rm IR}}\right) \frac{x+\xi}{(y+\xi)(y-x)} . 
\eeq

Adding these, we find, in the DGLAP region $x>\xi$, 
\begin{align}
&F_{h,{\rm DGLAP}} =\frac{\alpha_s C_F}{2\pi}  
\theta(x+\xi) \Biggl[ \delta(x-y) \left\{-\frac{1}{\epsilon_{\rm IR}^2}-\frac{1}{\epsilon_{\rm IR}}\left(2+\ln \frac{\mu^2}{-t}\right) -\frac{3}{2}\ln \frac{\mu^2}{-t}-\frac{1}{2}\ln^2\frac{\mu^2}{-t}\right\} \nn &\quad  + \left(\frac{1}{\epsilon_{\rm UV}}+\ln \frac{\mu^2}{-t}\right) \left\{\frac{x^2+y^2-2\xi^2}{y^2-\xi^2}\left[\frac{\theta(y-x)}{y-x}\right]_+ +\delta(x-y)\left(2+\ln \frac{(1-x)^2}{y^2-\xi^2}\right) \right\}\nn 
& \quad+\frac{(x^2+y^2-2\xi^2)\ln (y^2-\xi^2)-(x-y)^2}{y^2-\xi^2}\left[\frac{\theta(y-x)}{y-x}\right]_+ -2\frac{x^2+y^2-2\xi^2}{y^2-\xi^2}\left[\frac{\theta(y-x)\ln(y-x)}{y-x}\right]_+ \nn 
& \quad +\delta(y-x) \left(-\frac{1}{2}\ln^2\frac{y^2-\xi^2}{(1-x)^2}+\frac{\pi^2}{12}\right) \Biggr]. \label{hardtotal}
\end{align}
Note that the IR divergent  and Sudakov terms are all proportional to the leading order hard kernel  $K^{\rm LO} =\delta(x-y)$. In fact,  this is what the factorization formula  (\ref{f}) predicts.  We see that such a feature arises only after summing over graphs and does not hold for each graph individually, see the cancellation that took place between (\ref{dg}) and (\ref{sailh}).   (\ref{f}) also predicts that the leading double logarithms $(\alpha_s\ln^2t)^n$ at $n$-th order in perturbation theory are also proportional to $\delta(x-y)$, although subleading logarithms are in general not. As explained in \eqref{eq: repl 1} and below, the local structure $\delta(x-y)$ 
emerges after taking into account
the attachments of collinear (anti-collinear) gluons to the initial Wilson line $W(z_2\bar n,z_1\bar n)$ as well as to the anti-collinear (collinear) legs, by the eikonalization mechanism
\begin{align}
\frac{1}{(p_c + p_{\bar c})^2} \sim \frac{1}{\bar w \cdot p_c} \frac{1}{w \cdot p_{\bar c}}.
\end{align}
Both of the aforementioned contributions produce Wilson lines along the same direction so that they cancel at the junction at $z_2 \bar w$  ($z_1 w$)  producing the final  Wilson line $W_c^\dagger(z_1\bar w)$ ($W_{\bar c}(z_2w)$) in  \eqref{eq: repl 1} that protrudes from the collinear (anti-collinear) quark field. This will be  seen  in another example in the next subsection.

\subsection{Collinear region}
In the collinear region, again there is an important cancellation between the box and sail diagrams 
\begin{align}
 \tilde{F}_c^{\rm box}(zn/2,-zn/2,p_1,p_2) &= -2ig^2C_F\int \frac{d^dk}{(2\pi)^{d}} \frac{\bar u_{\bar c}(p_2) \Slash n_{\perp} u_c(p_1)}{k^2 (k+p_1)^2} \frac{\bar w \cdot (p_1+k)}{\bar w\cdot k} e^{-i\frac{z}{2}w\cdot p_2-i \frac{z}{2} \bar w\cdot p_1} e^{ - i z \bar w \cdot k},
\end{align}
where $u_c(p) = \frac{\s w \s{\bar w}}{4} u(p), \,u_{\bar c}(p) = \frac{\s{\bar w}\s w }{4} u(p) $ are the quark spinors projected onto the leading components.
\begin{align}
\tilde{F}_c^{\rm sail}(zn/2,-zn/2,p_1,p_2) &= - 2ig^2C_F \int \frac{d^dk}{ (2\pi)^{d}} \frac{\bar u_{\bar c}(p_2) \Slash n_{\perp} u_c(p_1)}{k^2 (p_1+k)^2} \frac{\bar w\cdot (p_1+k)}{\bar w \cdot k} e^{-i\frac{z}{2} w \cdot p_2 - i\frac{z}{2} \bar w \cdot p_1}  \left ( 1 - e^{- i z \bar w\cdot k} \right ).
\end{align}
The sum is 
\begin{align} \notag
\tilde{F}_c^{\rm box} + \tilde{F}_c^{\rm sail} &= -2ig^2C_F\int \frac{d^dk}{(2\pi)^{d}} \frac{\bar u_{\bar c}(p_2) \Slash n_{\perp} u_c(p_1)}{k^2 (k+p_1)^2} \frac{\bar w \cdot (p_1+k)}{\bar w \cdot k} e^{-i\frac{z}{2} w \cdot p_2 -i \frac{z}{2} \bar w \cdot p_1}  \left ( \cancel{ e^{ - i z \bar w k} } + 1 - \cancel{ e^{- i z \bar wk}  }\right )
\\
&=\frac{2g^2C_F}{(4\pi)^{d/2}} e^{-i\frac{z}{2} w \cdot p_2 -i \frac{z}{2} \bar w \cdot p_1} \bar u_{\bar c}(p_2) \Slash n_{\perp} u_c(p_1)  \frac{\Gamma(1+\epsilon)\Gamma(1-\epsilon)\Gamma(2-\epsilon)}{\epsilon^2 \Gamma(2-2\epsilon)} \left(\frac{\mu^2}{-p^2_1}\right)^{\epsilon}. \label{cancel}
\end{align}
Only after this cancellation does  the result become proportional to the delta function 
\begin{align}
F_c^{\rm box} + F_c^{\rm sail} &= \frac{\alpha_sC_F}{ 2\pi} 
\delta(y-x) \Bigg \{ \frac{1}{\epsilon_{\rm IR}^2} + \frac{1}{\epsilon_{\rm IR}} \left ( 1 + \ln \frac{\mu^2}{-p^2_1} \right ) + \frac{1}{2} \ln^2 \frac{\mu^2}{-p^2_1} + \ln \frac{\mu^2}{-p^2_1} - \frac{\pi^2}{12} + 2 \Bigg \}.
\end{align}
The anti-collinear contribution is simply obtained by $\ln(-p^2_1)\to \ln (-p_2^2)$. 
The self energy diagrams receive contributions only from the collinear and anti-collinear regions without any approximation 
\beq
F_c^{\rm self}+F_{\bar{c}}^{\rm self}= \frac{\alpha_sC_F}{2\pi} 
\left(-\frac{1}{4}\right)\delta(y-x)\left(\frac{1}{\epsilon_{\rm UV}}+\ln \frac{\mu^2}{-p_1^2}+1 \right) + (p^2_1\leftrightarrow p_2^2).
\eeq

\subsection{Ultrasoft region}
The ultrasoft contribution of the box diagram is 
\beq
 \tilde{F}_s^{\rm box} &=& -2ig^2C_F \bar u(p_2) \Slash n_{\perp} u(p_1) e^{-i\frac{z}{2}n \cdot p_2 - i\frac{z}{2}n \cdot p} \int \frac{d^dk}{(2\pi)^d} \frac{\bar w \cdot p_1 \, w \cdot p_2}{k^2 (p_1^2 + w \cdot k \, \bar w \cdot p_1 + i0) (p_2^2 + \bar w \cdot k \, w \cdot p_2 + i0)} \nn 
&=&\frac{2g^2C_F}{(4\pi)^{d/2}}\bar u(p_2) \Slash n_{\perp} u(p_1)e^{-i\frac{z}{2} n\cdot p_2 - i\frac{z}{2}n \cdot p}\frac{\Gamma^2(1+\epsilon)\Gamma(1-\epsilon)}{-\epsilon^2} \left(\frac{-t\mu^2}{p_1^2p_2^2}\right)^\epsilon  .
\eeq
After Fourier transform, this becomes 
\beq
F_s^{\rm box} = \frac{\alpha_s C_F}{2\pi} 
\delta(y-x) \Bigg \{ - \frac{1}{\epsilon^2_{\rm IR}} - \frac{1}{\epsilon_{\rm IR}} \ln \frac{-t\mu^2}{p^2_1 p_2^2} - \frac{1}{2} \ln^2 \frac{-t\mu^2}{p_1^2 p_2^2} - \frac{\pi^2}{4} \Bigg \}.
\eeq

\subsection{Factorization at one-loop}

Adding all the contributions, we find 
\begin{align}
&F_{{\rm DGLAP}} =\frac{\alpha_s C_F}{2\pi} 
\theta(x+\xi) \Biggl[ \frac{\delta(x-y)}{2} \biggl\{ -\ln^2\frac{\mu^2_F}{-t}-3\ln \frac{\mu^2_F}{-t}+ \ln^2\frac{\mu_F^2}{-p^2_1} + \ln^2 \frac{\mu_F^2}{-p_2^2} \nn &+ \frac{3}{2}\ln \frac{\mu_F^2}{-p^2_1}+\frac{3}{2}\ln \frac{\mu_F^2}{-p_2^2} -\ln^2 \frac{-t\mu^2_F}{p^2_1p_2^2} +7-\frac{2\pi^2}{3}-\ln^2\frac{y^2-\xi^2}{(1-x)^2} \biggr\} \nn &+ \left(\frac{1}{\epsilon_{\rm UV}}+\ln \frac{\mu^2_{\rm UV}}{-t}\right) \left\{\frac{x^2+y^2-2\xi^2}{y^2-\xi^2}\left[\frac{\theta(y-x)}{y-x}\right]_+ +\delta(x-y)\left(\frac{3}{2}+\ln \frac{(1-x)^2}{y^2-\xi^2}\right) \right\} \label{total} \\
&+\frac{(x^2+y^2-2\xi^2)\ln (y^2-\xi^2)-(x-y)^2}{y^2-\xi^2}\left[\frac{\theta(y-x)}{y-x}\right]_+ -2\frac{x^2+y^2-2\xi^2}{y^2-\xi^2}\left[\frac{\theta(y-x)\ln(y-x)}{y-x}\right]_+  \Biggr]. \nonumber 
\end{align}
Note that all the IR divergences $\frac{1}{\epsilon_{\rm IR}^2},\frac{1}{\epsilon_{\rm IR}}$ have canceled. 
Collecting  the logarithmic terms, we recover the well-known  Sudakov double and single logarithms for off-shell external particles \cite{Sudakov:1954sw}
\begin{align}
&\frac{1}{2}\left(-\ln^2\frac{\mu_F^2}{-t}+ \ln^2\frac{\mu^2_F}{-p_1^2} + \ln^2 \frac{\mu_F^2}{-p_2^2}-\ln^2 \frac{-t\mu_F^2}{p_1^2p_2^2} +\frac{3}{2}\ln \frac{t}{p_1^2}+\frac{3}{2}\ln \frac{t}{p_2^2} \right) \nn
&= -\ln \frac{t}{p_1^2}\ln \frac{t}{p_2^2} +\frac{3}{4}\left(\ln \frac{t}{p_1^2}+\frac{t}{p_2^2}\right).
\end{align} 
For the present purpose, however,  we formally keep $\mu_F$ and factorize  (\ref{total}) in the form (\ref{f}). 
 In the DGLAP region, 
\begin{align}
&K_{\rm DGLAP}(x,y,\xi,t,\mu_{\rm UV},\mu_F)= \delta (x-y) + \frac{\alpha_sC_F}{2\pi} \Biggl[\frac{\delta(x-y)}{2} \biggl\{ -\ln^2\frac{\mu^2_F}{-t}-3\ln \frac{\mu^2_F}{-t}  +\frac{\pi^2}{6}-\ln^2\frac{y^2-\xi^2}{(1-x)^2} \biggr\} \nn &\quad  + \left(\frac{1}{\epsilon_{\rm UV}}+\ln \frac{\mu^2_{\rm UV}}{-t}\right) \left\{\frac{x^2+y^2-2\xi^2}{y^2-\xi^2}\left[\frac{\theta(y-x)}{y-x}\right]_+ +\delta(x-y)\left(\frac{3}{2}+\ln \frac{(1-x)^2}{y^2-\xi^2}\right) \right\}\label{d} \\ 
&\quad +\frac{(x^2+y^2-2\xi^2)\ln (y^2-\xi^2)-(x-y)^2}{y^2-\xi^2}\left[\frac{\theta(y-x)}{y-x}\right]_+ -2\frac{x^2+y^2-2\xi^2}{y^2-\xi^2}\left[\frac{\theta(y-x)\ln(y-x)}{y-x}\right]_+  \Biggr].\nonumber
\end{align}
 The jet and soft functions are 
\beq
J(p^2,\mu_F)= 1+\frac{\alpha_sC_F}{8\pi}\left(2\ln^2 \frac{\mu_F^2}{-p^2} +3\ln \frac{\mu_F^2}{-p^2} +7-\frac{\pi^2}{3}\right) ,\label{j}
\eeq
\beq
S(t,p_1^2,p_2^2,\mu_F)=1-\frac{\alpha_sC_F}{4\pi}\left(\ln^2 \frac{-t\mu_F^2}{p_1^2p_2^2}+\frac{\pi^2}{2}\right). \label{s}
\eeq
 The non-logarithmic terms in $J,S$ have been fixed to match their operator definitions (\ref{jdef}) and (\ref{sdef}). 
From (\ref{j}) and (\ref{s})
\beq
\frac{d}{d\ln \mu} JJS(\mu) = \Gamma_{\rm cusp}\left(-\ln \frac{-t}{\mu^2}+\frac{3}{2}\right)JJS(\mu) .\label{jjs}
\eeq
The solution of this equation is given by (\ref{rg}).

In the ERBL region $x<\xi<y$, we again find that $\frac{1}{\epsilon_{\rm IR}}$ poles cancel 
 \begin{align}
K_{\rm ERBL}(x,y,\xi,t,\mu_{\rm UV})= \frac{\alpha_s C_F}{2\pi}\Biggl[\left(\frac{1}{\epsilon_{\rm UV}}+\ln \frac{\mu_{\rm UV}^2}{-t}\right)\frac{x+\xi}{y+\xi}\left(\frac{1}{2\xi}+\frac{1}{y-x}\right) -\frac{x+\xi}{2\xi(y+\xi)}
 \nn 
  + \frac{1}{2\xi (y-x)(y^2-\xi^2)}\Biggl\{ (x-y)(x+\xi)(y+\xi)\ln \frac{\xi^2-x^2}{4\xi^2}  +2\xi x^2 \ln \frac{2\xi(y+\xi)}{(x+\xi)(y-x)}  \nn 
  + 2\xi^3\ln \frac{\xi+x}{\xi-x}+2\xi y^2 \ln \frac{(y+\xi)(\xi-x)}{2\xi(y-x)}+4\xi^3\ln \frac{y-x}{y+\xi}\Biggr\}\Biggr].  \label{der}
\end{align}
Note that this result entirely comes from the hard region $k^\mu \sim \sqrt{-t}$. The other regions do not contribute at all. 
Since the tree level amplitude $\delta(x-y)$ vanishes, the Sudakov logarithms are not visible at this order. But they will be present starting from two-loops, as predicted by (\ref{f}). 

Before leaving this section, we should mention  that the result (\ref{d}) has been derived under the assumption that it is to be convoluted in $y$ with a smooth test function $g(y)$, see (\ref{testf}). An important exception, which will be encountered later, is the case where  $g(y)$ is the  delta function $\delta(1-y)$. Naively,  (\ref{d}) leads to an ill-defined expression $\delta(1-x)\ln (1-x)$. It is easy to check that, in this case, one just needs  to remove the $\delta(x-y)\ln(1-x)$ terms in (\ref{d}) (two occurrences), set $y=1$, and regard the plus prescription $\frac{1}{[1-x]_+}=\frac{1}{1-x}-\delta(1-x)\int_0^1 \frac{dx'}{1-x'}$ as a distribution in $x$ in the ordinary sense. In this way, the result in \cite{Bhattacharya:2023wvy} is recovered. For a later purpose, however, it is more convenient to use a modified prescription 
\beq
\left[\frac{f(x)}{1-x}\right]^\xi_+\equiv \frac{1}{1-x}-\delta(1-x)\int_\xi^1 \frac{f(x')dx'}{1-x'}.
\eeq
We can then write, at $y=1$,  
\begin{align} \notag
&K(x,\xi,t) = \delta(1-x) +  \frac{\alpha_sC_F}{2\pi}\theta(x-\xi) \Biggl[\frac{\delta(1-x)}{2} \biggl\{ -\ln^2\frac{\mu^2_F}{-t}-3\ln \frac{\mu^2_F}{-t}  +\frac{\pi^2}{6}-\ln^2\frac{1+\xi}{1-\xi} \biggr\} \label{Ky=1} \\ &\quad  + \ln \frac{\mu^2_{\rm UV}}{-t} \left\{\frac{1+x^2-2\xi^2}{1-\xi^2}\left[\frac{1}{1-x}\right]^\xi_+ +\delta(1-x)\left(\frac{3}{2}-\ln \frac{1+\xi}{1-\xi}\right) \right\} \nn
&\quad +\frac{(1+x^2-2\xi^2)\ln (1-\xi^2)-(1-x)^2}{1-\xi^2}\left[\frac{1}{1-x}\right]^\xi_+ -2\frac{1+x^2-2\xi^2}{1-\xi^2}\left[\frac{\ln(1-x)}{1-x}\right]^\xi_+  \Biggr]
\nn
&\quad + \frac{\alpha_s C_F}{2\pi}  \theta(|\xi| - x) \Bigg \{  \ln  \frac{\mu^2_{\rm UV}}{-t}  \frac{x+\xi}{1+\xi}\left(\frac{1}{2\xi}+\frac{1}{1-x}\right) -\frac{x+\xi}{2\xi(1+\xi)}
\\ \notag
&\qquad + \frac{1}{2\xi (1-x)(1-\xi^2)}\Bigg ( (1-x)(x+\xi)(1+\xi)\ln \left ( \frac{4\xi^2}{\xi^2-x^2} \right )
 +2\xi x^2 \ln \frac{2\xi(1+\xi)}{(x+\xi)(1-x)} \\ \notag
&\qquad + 2\xi^3\ln \left ( \frac{\xi+x}{\xi-x} \right ) +2\xi \ln \left ( \frac{(1+\xi)(\xi-x)}{2\xi(1-x)} \right )+4\xi^3\ln \left ( \frac{1-x}{1+\xi} \right ) \Bigg ) \Bigg \}
+ \f O(\alpha_s^2).
\end{align}
The region $-1<x<-\xi$ is absent since we do not consider an anti-quark in the  initial state.  One can check that $K(x,\xi)$ is continuous at $x=\xi$ and vanishes at $x=-\xi$.  
Integrating (\ref{Ky=1}) over $x$, we recover the result \eqref{eq: local hard} independently of the value of $\xi$. The $\xi$-dependence cancels between the DGLAP and ERBL contributions.

\section{From quark GPD to proton GPD} \label{proton new}

So far, our discussion has been confined to the unphysical case of the quark-in-quark GPD (\ref{def}). In this  section, we explore the implications of our result on the large-$t$ behavior of the quark-in-proton GPD (i.e., ordinary proton GPD) 
\beq
\bar n \cdot P\int \frac{dz}{2\pi}e^{ixzP\cdot \bar n}\bra{P_2} \bar \psi(-z \bar n/2) \sbar n W 
\psi(z \bar n/2) \ket{P_1}=H_q(x,\xi,t) \bar N(P_2)\sbar n N(P_1) +\cdots,
\label{nucleongpd}
\eeq
where $P^\mu=\frac{P^\mu_1+P^\mu_2}{2}$ and $N$ is the nucleon spinor and now $t = (P_2 - P_1)^2$.   
The nucleon spin-flip GPD $E_q$ (omitted in (\ref{nucleongpd})) is power-suppressed at large-$t$ and requires a separate treatment \cite{Tong:2022zax}. We will not  discuss it in this paper except for a brief mention in Sec.~\ref{erblregion}. 
As explained in the introduction, the large-$t$ behavior of the electromagnetic form factors of  hadrons is a subject of considerable debate. Analogous  discussions for GPDs are few and far between  \cite{Diehl:1999ek,Hoodbhoy:2003uu,Bhattacharya:2023wvy}.

(\ref{nucleongpd}) is usually defined in  frames  where $P^\mu_{1,2}\sim n^\mu$ are collinear and satisfy  
\beq
\bar n \cdot P_1 = (1+\xi)\bar n \cdot P, \qquad \bar n\cdot P_2 = (1-\xi)\bar n \cdot P. \label{1+xi}
\eeq
Comparing with (\ref{eq: nP}), we see that the variable $y$ (more precisely, $\frac{y\pm \xi}{1\pm \xi}$)  can be interpreted as the momentum fraction of the proton carried by the `active' quark. As in Sec.~\ref{quarkin}, 
we transform to the $(w,\bar w)$ frame (see Appendix \ref{app A}, set $y=1$ there) in which the proton momenta take the form
\begin{align}
P_1^{\mu} = (1+\xi) \bar n \cdot P \frac{w^{\mu}}{2}, \qquad P_2^{\mu} = (1-\xi) \bar n \cdot P \frac{\bar w^{\mu}}{2},
\end{align}
where we neglected the proton mass.  The factorization of $H_q$ can be  most conveniently done in the $(w,\bar w)$ frame. As mentioned in the Introduction, there are two contributions to $H_q$: (i) The standard   Efremov-Radyushkin-Brodsky-Lepage factorization \cite{Efremov:1979qk,Lepage:1980fj},  which we simply refer to as the  `hard scattering'  contribution in the following. (ii) The Feynman contribution, which is the main interest of this paper.    
\begin{figure}[t]
\vspace{-20mm}
    \centering
\includegraphics[width=1\linewidth]{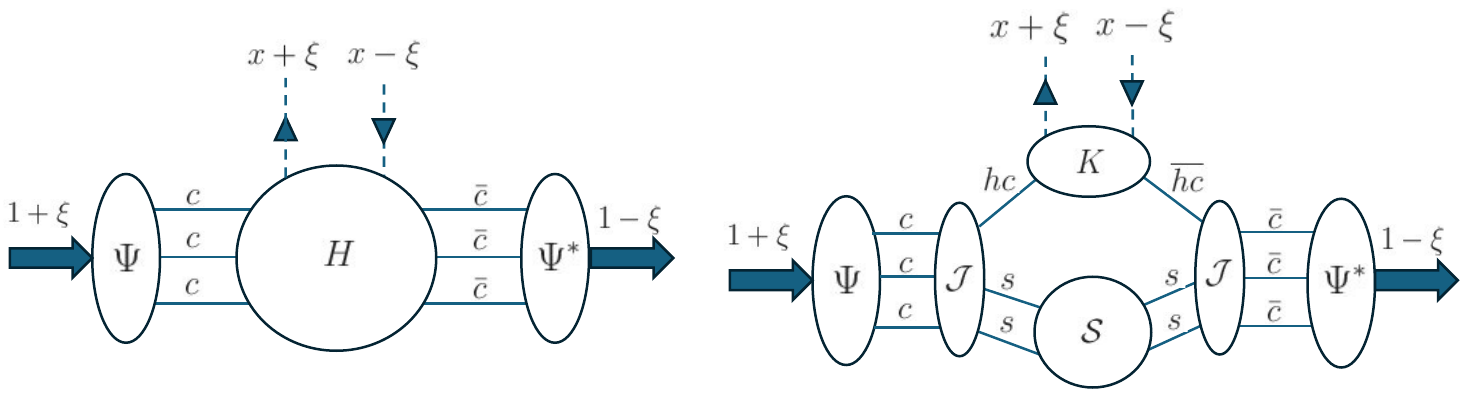}
\vspace{-65mm}
    \caption{Left: The hard scattering contribution to GPDs that dominates in the asymptotic regime $|t|\to \infty$.  Right: The Feynman  contribution in the asymptotic regime. The labels $hc$, $c$ and $s$ on quark lines  denote `hard-collinear', `collinear' and `soft'.  }
    \label{s2}
\end{figure}
The former  schematically reads \cite{Hoodbhoy:2003uu}
\begin{align} \label{eq: hard scattering}
H_q^{\text{hard}} = {\rm H} \otimes \Psi \otimes \Psi^* +{\cal O}\left(\frac{1}{t^3}\right),
\end{align}
where $\Psi(x_i)$ is the proton distribution amplitude (DA) and the symbol $\otimes$ denotes convolution integrals over the momentum fractions $x_{i=1,2,3}$. The hard kernel ${\rm H}$ is calculable in perturbation theory and starts at order $\alpha_s(\sqrt{-t})^2$.  The corresponding reduced graph is shown in Fig.~\ref{s2} (left). In the present context, it is important to mention that Sudakov double logarithms are absent in the hard scattering contribution because they cancel due to the color neutrality of the proton.

On the other hand, Sudakov double logarithms do not cancel in the Feynman contribution $H_q^{\rm Feyn}$ due to the presence of soft spectators.  The QCD factorization of the Feynman contribution to the electromagnetic form factor $F_{1}^{\rm Feyn}$ has been established by Kivel and Vanderhaeghen  \cite{Kivel:2010ns,Kivel:2012mf,Kivel:2013sya} using SCET. Using the result of Sec.~\ref{quarkin}, we will generalize their approach to the GPD  $H_q^{\rm Feyn}$.

For this purpose, we first need to specify the quark virtuality $\mu_{hc}\equiv \sqrt{-p^2}$ which remained unspecified in Sec.~\ref{quarkin}. 
It should be given by 
\beq
-p^2 = \mu_{hc}^2 = \sqrt{-t} \Lambda \gg \Lambda^2,\label{new}
\eeq
where $\Lambda = {\cal O}(\Lambda_{\rm QCD})$ is a nonperturbative scale.  
Physically, the emergence of this `hard-collinear' scale can be understood by noting that spectators in the Feynman contribution are soft $p_s^2 \sim(P_1-p_1)^2\sim \Lambda^2$. This gives  $w \cdot p_1 \sim -\frac{\Lambda^2}{(1-y)\bar w \cdot P_1}$ and therefore,
\beq
p^2_1\sim -\frac{\Lambda^2}{1-y} \sim \sqrt{-t}\Lambda, 
\eeq
where $1-y\sim \frac{\bar w \cdot p_s}{\bar w \cdot P_1}\sim \frac{\Lambda}{\sqrt{-t}}$. 
In the context of perturbative QCD,  the scale (\ref{new}) was first noticed  in the analysis of certain two-loop (i.e., order $\alpha_s^4$) diagrams that topologically resemble the Feynman contribution  \cite{Duncan:1979hi}. It is connected to end-point divergences due to the soft-collinear overlap region in the hard scattering contribution. Later, the hard-collinear  scale  and the associated power counting  parameter 
\beq
\lambda = \frac{\mu_{hc}}{\sqrt{-t}} =  \frac{\sqrt{\Lambda}}{(-t)^{1/4}},
\label{lambdahc}
\eeq
were used in the proof of the factorization of $F^{\rm Feyn}_{1}$ in  \cite{Kivel:2010ns}.   
 Accordingly, we switch to the nomenclature in \cite{Kivel:2010ns} and rename
\beq
p^\mu_{hc} = \sqrt{-t}(1,\lambda^2, \lambda), \qquad p^\mu_{\widebar{hc}} =\sqrt{-t}(\lambda^2,1,\lambda),  \label{hcmode}
\eeq
in the $(w,\bar w)$ frame 
as the hard-collinear and hard-anti-collinear modes, and newly introduce the `collinear', `anti-collinear' and `soft' modes 
\beq
&&p^\mu_c\sim \sqrt{-t}(1,\lambda^4,\lambda^2)=\left(\sqrt{-t}, \frac{\Lambda^2}{\sqrt{-t}},\Lambda \right), \qquad p^\mu_{\bar c}\sim \sqrt{-t}(\lambda^4,1,\lambda^2)=\left( \frac{\Lambda^2}{\sqrt{-t}},\sqrt{-t},\Lambda\right),\nn
 &&\qquad  \qquad \qquad \qquad \qquad p_s^{\mu}\sim \sqrt{-t}(\lambda^2,\lambda^2,\lambda^2)=(\Lambda, \Lambda,\Lambda).
 \label{soft2}
\eeq
Note that the `soft' mode was called  the ultrasoft mode (\ref{region}) in Section 2.

The reduced graph for the Feynman contribution, called the `soft rescattering' contribution in \cite{Kivel:2010ns}, is shown in Fig.~\ref{s2} (right). Schematically
\begin{align} \label{eq: soft rescattering 2}
H_q^{\text{Feyn}} = K \otimes {\cal S} \otimes {\cal J} \otimes {\cal J^*} \otimes \Psi \otimes \Psi^* +\cdots \, .
\end{align}
$K$ is the kernel calculated in Section \ref{sec: 1loop calc}, $\Psi$ is the proton DA, ${\cal S}$ is a soft function describing the soft modes $\sim p_s$, and the jet function ${\cal J}$ describes the hard-collinear interactions (hard-anti-collinear for ${\cal J}^*$) at the scale $\mu_{hc}$. For their    precise definitions in SCET and the meaning of the convolutions in (\ref{eq: soft rescattering 2}), we refer to \cite{Kivel:2010ns}.

The factorization formula  \eqref{eq: soft rescattering 2} fully separates  physics according to the two-fold hierarchy of scales $\Lambda \ll  \mu_{hc} \ll  \sqrt{-t}$. $K$ arises from integrating out modes with virtuality $-t$ and ${\cal J}, {\cal  J}^*$ arise from integrating out modes with virtuality $\mu_{hc}^2$. It is therefore necessary to consider the  following two cases  
\begin{itemize}
    \item[(i)] The asymptotic regime 
\beq
10\sim 25\ {\rm  GeV}^2 \ll |t|\to \infty, \label{re2}
\eeq 
where $\mu^2_{hc}= \sqrt{-t}\Lambda\gg 1$ GeV$^2$. This regime is mainly  of theoretical interest. 
\item[(ii)] The pre-asymptotic regime 
\beq
\Lambda^2\ll |t|\lesssim 10\sim 25\ {\rm  GeV}^2, \label{re1}
\eeq
where $\mu_{hc}^2\lesssim 1$ GeV$^2$. In practice, this  covers almost the entire kinematically  accessible region in experiments. 
There is a large uncertainty in the boundary value $|t_{th}|=10\sim 25$ GeV$^2$. It is only a rough estimate from the condition $\sqrt{-t_{th}}\Lambda \sim 1$ GeV$^2$ with  $200<\Lambda< 300$ MeV. 
\end{itemize}
Since the ${\cal J}$ function is given by a series in $\alpha_s(\mu_{hc})$, the factorization in \eqref{eq: soft rescattering 2} is only useful in the asymptotic regime, but not in the pre-asymptotic regime. In the former case, \eqref{eq: soft rescattering 2} is actually a leading twist contribution, but higher order in $\alpha_s$, $H_q^{\rm Feyn}\sim \alpha_s^4/t^2$  to be compared to \eqref{eq: hard scattering} $H_q^{\rm hard}\sim \alpha_s^2/t^2$. 
In the latter case, we might as well factorize only the physics at the hard scale $\sqrt{-t}$ 
and describe everything at and below $\mu_{hc}$ by a nonperturbative function, which we will denote by $h_q$. We then have, in the pre-asymptotic regime, 
\begin{align} 
H_q^{\text{Feyn}}(x,\xi,t,\mu) =  \int dy \, K(x,y,\xi, t,\mu,\mu_{hc}) h_q(y,\xi,t,\mu_{hc}), \label{fac2}
\end{align}
where $K$ is the same as before and $\mu\sim \sqrt{-t}$. We remind the reader that $y$ is the momentum fraction of the active quark with respect to the parent proton. 
In Section \ref{sec: SCET soft}, we will give the operator definition of $h_q$ and show that in the SCET formalism 
\begin{align}
h_q(y,\xi,t,\mu_{hc}) = f^q_1(t,\mu_{hc}) \delta(1-y), \label{delta}
\end{align}
so that \eqref{fac2} becomes a simple product. In fact, the same remark applies to (\ref{eq: soft rescattering 2}) as well, namely, the first convolution with $K$ is actually a simple product. Indeed, this makes sense physically, since kinematically the active quark must carry most of the momentum of the parent proton $1-y \sim \Lambda/\sqrt{-t}$. In the leading power approximation, $\Lambda$ is neglected with respect to $\sqrt{-t}$, which effectively sets $y = 1$. Since $f_1^q$ is nonperturbative in the pre-asymptotic region, (\ref{fac2}) counts as order $\alpha_s^0$. It is therefore expected to dominate over the hard scattering contribution $H_q^{\rm hard}\sim \alpha_s^2(\sqrt{-t})/t^2$ in the entire pre-asymptotic regime. 

The function $f_1^q(t)$ was first defined in \cite{Kivel:2010ns}, and it is the sole non-perturbative input that appears in the leading power Feynman (soft rescattering) contribution to the electromagnetic form factor $F_{1q}$ as well as the GPD $H_q$. (Note that it does not depend on $\xi$.)  
The multiplicative factorization is a great simplification and we will explore its implications in the next section. 
On the other hand, (\ref{delta}) leads to an extreme prediction $H_q^{\rm Feyn}(x,\xi,t) \approx \left( \delta(1-x) + \f O(\alpha_s(\sqrt{-t})) \right) f_1^q(t)$ that the leading term of the GPD is the delta function at $x=1$. In Section~\ref{lc}, we offer an alternative `hybrid' formulation where we retain the convolution structure \eqref{fac2} and model $h_q$ as a function of $y$ using light-cone wave functions.

\section{SCET formulation of the Feynman contribution} \label{sec: SCET soft}

In this section we present a SCET analysis of the function $f_1(t)$ in (\ref{delta}) and provide a precise operator definition. We will also show how to obtain its power counting in $\lambda$, from which the $t$-dependence can be inferred. We start by recalling \eqref{eq: factorization 1}, but now we only integrate out modes with virtuality   down to $\mu_{hc}^2 = \Lambda \sqrt{-t}$. Let us denote the resulting effective field theory (EFT) by SCET($C,\bar C,s$), where $C (\bar C)$ denotes the combination of hard-(anti-)collinear modes
\begin{align}
\xi_C = \xi_c + \xi_{hc}, \qquad \xi_{\bar C} = \xi_{\bar c} + \xi_{\widebar{hc}},
\end{align}
etc. The ultrasoft Wilson lines $Y_{\bar w}^\dagger Y_w$ are to be reinterpreted as soft ($s$) Wilson lines according to   (\ref{soft2}). Furthermore, we consider contributions due to anti-quarks and gluons, which are naively of the same power in $\lambda$ at the naive SCET($C,\bar C,s$) power counting. Then the expansion of the non-local current reads
\begin{align}  \notag
\bar \psi^q(z_2\bar n) \sbar n W(z_2\bar n,z_1\bar n) \psi^q(z_1 \bar n)
&= \sum_{q'} \int ds_1 ds_2 \, \w K_{qq'}(s_1 , s_2,z_1-z_2) \, O^{q'}(z_2+ s_2, z_1 +s_1) 
\\  \label{eq: factorization 3}
&+ \int ds_1  ds_2 \, \w K_{qg}(s_1 , s_2,z_1-z_2) \, O^{g}(z_2+ s_2, z_1 +s_1) + ``\f O(\lambda^3)",
\end{align}
where
\begin{align}
O^q(z_2,z_1) &= \bar \chi_{\bar C}^q(z_2 w) \sbar n_{\perp} Y_{\bar w}^{\dagger}\left ( 0\right ) Y_{w}\left ( 0 \right ) \chi_{C}^q(z_1 \bar w), \label{antiq}
\\ \label{eq: Og}
O^g(z_2,z_1) &= \f A_{\bar C \perp}^{\mu A}(z_2 w) [\f Y_{\bar w}^{\dagger}(0) \f Y_w(0)]^{AB} \f A_{C \perp \mu}^B(z_1 \bar w) + (z_1 \leftrightarrow z_2).
\end{align} 
The gluon SCET building blocks are defined in the usual way
\begin{align}
\f A_{C}^{\mu} = W_{C}^{\dagger} [(i\partial^{\mu} + g A_{C}^{\mu}) W_{C}],
\end{align}
and analogously for $\f A_{\bar C}^{\mu}$. $\f Y_w$ denotes the soft Wilson line in the adjoint representation. Note that the sum $\sum_{q'}$ in \eqref{eq: factorization 3} also goes over anti-quark contributions $O^{\bar q}(z_2,z_1) = O^q(z_1,z_2)^{\dagger}$. $\widetilde{K}_{qg}$ (in momentum space) has been calculated in \cite{Bhattacharya:2023wvy} to one-loop. 

Correspondingly, we  denote the EFT that only contains $c,\bar c,s$ by SCET($c,\bar c,s$). \eqref{eq: factorization 3} describes the matching of QCD onto SCET($C,\bar C,s$). Since we are interested in the pre-asymptotic regime where $\mu_{hc}$ is not quite perturbative, we  do not  integrate out the hard-collinear modes. Instead, they are part of non-perturbative matrix elements $\langle P_2|O|P_1\rangle$ in SCET($C,\bar C,s$). However, unlike in SCET($c,\bar c,s$), the power counting in SCET($C,\bar C,s$) is complicated because  $hc,\widebar{hc}$ fields do not scale  homogeneously in $\lambda$. Yet, we can estimate  the $\lambda$-scaling of various matrix elements by considering how SCET($C,\bar C,s$) operators match onto SCET($c,\bar c,s$) operators and then using the SCET($c,\bar c,s$) counting rules.

Let us  present an argument showing that, when taking the matrix element $\bra{P_2} \cdots \ket {P_1}$ of \eqref{eq: factorization 3}, the anti-quark and gluon contributions are power-suppressed. 
We  first review a simplified version of the argument from \cite{Kivel:2010ns} and show that the Feynman contribution has the same power as the hard scattering contribution. The latter can be estimated as  
\begin{align}
\bra{P_2} \bar \xi_{\bar c} \bar \xi_{\bar c}  \bar \xi_{\bar c} \xi_c \xi_c \xi_c \ket{P_1} \sim \lambda^8 \sim 1/t^2, \label{8}
\end{align}
since the proton state can only overlap with a minimum of three collinear quark operators. 
We remark that the six-quark operator is part of the ``$\f O(\lambda^3)$''  terms in \eqref{eq: factorization 3} because it is power-suppressed in the naive SCET($C,\bar C,s$) operator counting. However it turns out to be a leading power contribution at the matrix element level.

\begin{figure}
    \centering
\includegraphics[width=1\linewidth]{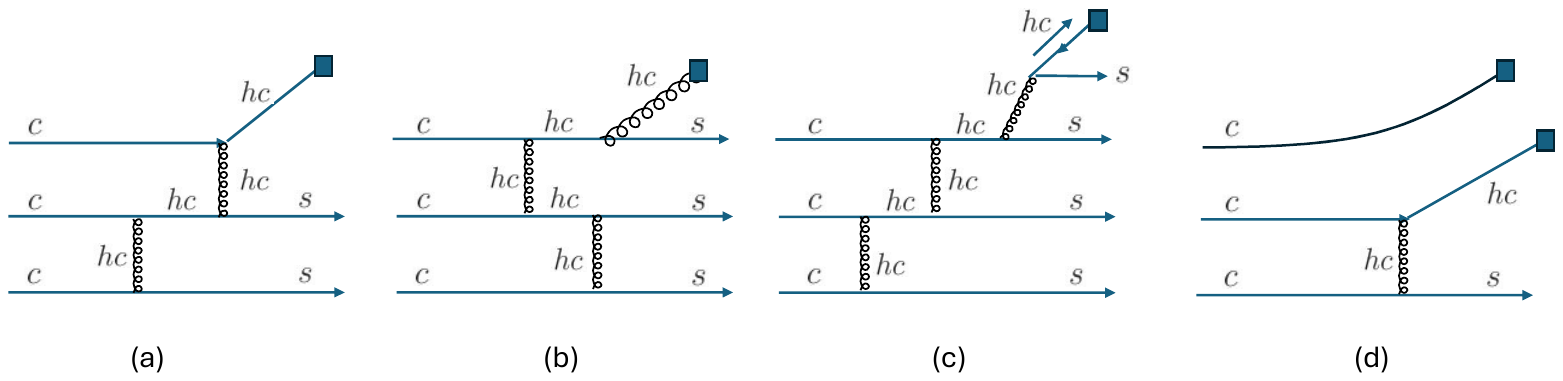}
 \vspace{-7cm}
    \caption{Sample tree-level graphs for the SCET$(C,\bar C, s)$ $\rightarrow$ SCET$(c,\bar c,s)$ matching for operators (a) $\xi_{hc}$, (b) $A_{hc\perp}$ and (c) $\bar \xi_{hc}$.
    }
    \label{fig: hcmatching}
\end{figure}

As shown in \cite{Kivel:2010ns}, the bilinear operator  $O^q \simeq \bar \xi_{\bar C} \xi_{C}$  can produce the required six-quark operator $\bar \xi_{\bar c} \bar \xi_{\bar c}  \bar \xi_{\bar c} \xi_c \xi_c \xi_c$ at the same power through insertions of SCET($C,\bar C,s$) interactions. Hard-(anti-)collinear  modes can only interact through soft modes, so it is sufficient to consider the matching $C \rightarrow c + s$, or more specifically,  
\begin{align}
\xi_{C} \ket {P_1} \rightarrow {\cal J}_q \, \bar q_s \bar q_s \xi_c \xi_c \xi_c \ket {P_1}, \label{56}
\end{align}
and similarly for $\bar C \rightarrow \bar c + s$. Here $\f J_q$ is the matching coefficient function resulting from integrating out the hard-collinear modes. In position space, it may involve inverse soft derivatives $(w \cdot \partial )^{-1}, (\bar w \cdot \partial)^{-1} \sim \lambda^{-2}$, resulting from the hard-collinear propagators $\frac{1}{(p_c + p_s)^2} \sim \frac{1}{\bar w \cdot p_c \, w \cdot p_s}$, which makes the power-counting non-trivial. 
In \eqref{56} two soft anti-quark fields have to be present by momentum and baryon number conservation. A sample diagram that contributes to ${\cal J}_q$ is shown in Fig.~\ref{fig: hcmatching}(a). 
By examining the graph it is straightforward to see that the hard-collinear matching coefficient function must scale as $\f J_q \sim \lambda^{-6}$. To explain this, first note that there are four hard-collinear propagator denominators present in the diagram, giving a total of $(\lambda^{-2})^4 = \lambda^{-8}$.  To determine the scaling of the  numerator, 
we pick the $\gamma^{\nu} \rightarrow w^{\nu} \frac{\s{\bar w}}{2}$ component of the uppermost vertex, and correspondingly  take the large momentum component $p_{hc} \rightarrow \bar w \cdot p_{hc} \frac{\s{ w}}{2}\sim \lambda^0$ in the numerator of the adjacent  hard-collinear propagator. 
We are then forced to take the small $ w \cdot p_{hc} \frac{\s{\bar w}}{2} \sim \lambda^2$ component in the numerator of the hard-collinear propagator in the middle quark line, giving  ${\cal J}_q\sim \lambda^{-8}\times \lambda^2=\lambda^{-6}$. To keep the main presentation simple, we only perform this simple tree-level analysis here. A more general analysis must include all possible tree-level graphs and arbitrary hard-collinear loops. In Appendix \ref{app: pc all orders} we argue that the scaling $\f J_q \sim \lambda^{-6}$ must hold for all graphs and to all orders by Lorentz invariance and dimensional analysis. We also perform this analysis for the remaining cases discussed in this section in Appendix \ref{app: pc all orders}. We note however, that a more rigorous proof can be made along the lines of \cite{Beneke:2003pa}, which is beyond the scope of this work. 

The exact same considerations hold for the hard-anti-collinear sector. Furthermore, the soft quark fields scale as $(q_s \bar q_s )^2 \sim (\lambda^6)^2$. Finally, the purely collinear piece scales as $\xi_c \xi_c \xi_c \ket {P_1} \sim \lambda^4$ and the same for the purely anti-collinear piece. 
Therefore we obtain the estimate
\begin{align}
\bra{P_2} O^q \ket {P_1} \sim (\lambda^{-6})^2 \times (\lambda^6)^2 \times (\lambda^4)^2 = \lambda^8\sim 1/t^2,
\end{align}
which is the same power as the hard scattering contribution (\ref{8}). 

Next consider the gluon contribution. In order to produce a hard-collinear gluon, we need to have three soft anti-quark fields to conserve the baryon number. We thus consider the matching 
\begin{align}
A_{C} \ket {P_1} \rightarrow {\cal J}_g \bar q_s \bar q_s \bar q_s \xi_c \xi_c \xi_c \ket {P_1}.
\end{align}
A corresponding leading power and tree-level graph is shown in Fig.~\ref{fig: hcmatching}(b). It is straightforward to find that $\f J_g \sim \lambda^{-8}$. Together with  $(q_s \bar q_s)^3 \sim (\lambda^6)^3$, this gives 
\begin{align} \label{eq: Og me}
\bra {P_2} O^g \ket {P_1} \sim (\lambda^{-8})^2 \times (\lambda^6)^3 \times (\lambda^4)^2 = \lambda^{10} \sim 1/(-t)^{5/2}.
\end{align}
The sample diagram for the anti-quark contribution is shown in Fig.~\ref{fig: hcmatching}(c). The matching reads
\begin{align}
\bar \xi_{C} \ket{P_1} \rightarrow \f J_{\bar q} \bar q_s \bar q_s \bar q_s \bar q_s \xi_c \xi_c \xi_c \ket {P_1},
\end{align}
and we find that $\f J_{\bar q} \sim \lambda^{-10}$. Noting that we now have $(q_s \bar q_s)^4 \sim (\lambda^6)^4$, we obtain
\begin{align}
\bra{P_2} O^{\bar q} \ket {P_1} \sim (\lambda^{-10})^2 \times (\lambda^6)^4 \times (\lambda^4)^2 = \lambda^{12} \sim 1/t^3.
\end{align}
We see that each additional soft spectator quark leads to an additional power suppression by $1/\sqrt{-t}$. This is consistent with the well-known suppression of higher Fock states \cite{Lepage:1980fj}.

For completeness, we also consider the case where there is only a single soft spectator quark as in Fig. \ref{fig: hcmatching}(d) and show that it is power suppressed. This corresponds to the operator $\bar \xi_{\bar C} \bar \xi_{\bar C} \xi_{C} \xi_C$, which is also neglected in  \eqref{eq: factorization 3}. 
Again, we focus on the $C \rightarrow c + s$ part of the matching, which reads
\begin{align} \label{eq: xihc xic}
\xi_C \xi_C \ket {P_1} \rightarrow \f J'_{q} \, \bar q_s \xi_c \xi_c \xi_c \ket{P_1}.
\end{align}
Examining the graph in Fig. \ref{fig: hcmatching}(d), we see that $\f J'_{q} \sim \lambda^{-4} \times \lambda^2 = \lambda^{-2}$, where $\lambda^{-4}$ comes from the denominators and $p_{hc\perp}\sim \lambda^2$ from the numerator. Note that here it is important that the hard-collinear quark momentum $p_{hc}$ in Fig. \ref{fig: hcmatching}(d) has a scaling $p_{hc} \sim (1,\lambda^2, \lambda^2) \sqrt{-t}$ by momentum conservation.\footnote{It is a known complication that the hard-collinear scaling in tree-level graphs $p_{hc} \sim (1,\lambda^2, \lambda^2) \sqrt{-t}$ differs from its `on-shell' version \eqref{hcmode}  used for hard-collinear loop momenta.}  Hence
\begin{align}
\bra{P_2} \bar \xi_{\bar C} \bar \xi_{\bar C} \xi_{C} \xi_C \ket{P_1} \sim (\lambda^{-2})^2 \times (\lambda^6)^1 \times (\lambda^4)^2 = \lambda^{10} \sim 1/(-t)^{5/2},
\end{align}
namely, it is of the same power as the gluon contribution in \eqref{eq: Og me}.

We have thus argued that
\begin{align}   \notag
\bra{P_2} \bar \psi^q(z_2\bar n) \sbar n W(z_2\bar n,z_1\bar n) \psi^q(z_1 \bar n) \ket{P_1}^{\text{Feyn}}
&= \sum_{q'} \int ds_1 ds_2 \, \w K_{qq'}(s_1 , s_2,z_1-z_2) 
\\ \label{eq: factorization 4}
&\quad \times \bra{P_2} O^{q'}(z_2+ s_2, z_1 +s_1)  \ket{P_1} + \f O(\lambda^{10}).
\end{align}
Note that the flavor off-diagonal  matching kernels $\w K_{qq'}$ with $q\neq q'$ ($\w K_{ud}$ and $\w K_{du}$ in the proton case) start at two-loop $\alpha^2_s(\sqrt{-t})$, which is of the same size $\alpha^2_s(\sqrt{-t})$ as the hard scattering contribution  neglected in \eqref{eq: factorization 4}. We will therefore keep only the diagonal term $q'=q$ in the following.

At leading power, the collinear momentum operator in SCET$(C,\bar C,s)$, $\hat P_{C}$, commutes with $\hat P_{\bar C}$ and $\bar C$ fields, and annihilates the anti-collinear proton state (and vice-versa for $\hat P_{\bar C}$).  $\hat P_{C}$ and $\hat P_{\bar C}$ also commute with soft operators at leading power, so we have
\begin{align}
\bra{P_2} O^q(z_2,z_1) \ket {P_1} &= e^{-iz_1 \bar w\cdot P_1 + iz_2 w\cdot P_2} \bra{P_2} O^q(0,0) \ket {P_1}  + \f O(\lambda^{10}). \label{leadingpo} 
\end{align}
 We refer to the order $\lambda^{10}$ corrections in this equation as `kinematical' power corrections. These result from the non-commutativity of the hard-collinear momentum operators with soft fields beyond leading power. More precisely, the corresponding operators contain additional derivatives $w\cdot \partial$ or $\bar w\cdot \partial$ acting on soft fields. Consider for example the matrix element $\bra{p_s} \chi_{C}(z_1 \bar w) \ket{p_c}$,
where $\ket{p_c} (\ket{p_s})$ is a one- or multiparticle collinear (soft) state. This amplitude may appear as a sub-amplitude of $\bra{P_2} O^q \ket{P_1}$, see e.g. the graphs in Fig. \ref{fig: hcmatching}, where $p_s$ is the total momentum of the intermediate soft state. To all orders in $\lambda$ we have
\begin{align}
\bra{p_s} \chi_{C}(z_1 \bar w) \ket{p_c} = e^{-iz_1\bar w(p_c - p_s)} \bra{p_s} \chi_{C}(0) \ket{p_c}.
\end{align}
The exponential may be expanded as $e^{-iz_1\bar w\cdot(p_c - p_s)} = e^{-iz_1\bar w \cdot p_c} (1 + iz_1\bar w\cdot p_s + \f O(\lambda^4))$. In terms of interpolating operators, $\bar w \cdot p_s\sim \lambda^2$ can be written as a derivative $\bar w\cdot \partial$ acting on some combination of soft fields, say on the quark fields in  \eqref{56}. Note that in this argument it is critical that $z_1 \sim \lambda^0$, since the large $\sim \lambda^0$ component of the collinear momentum flows out into the hard scattering through the vertex at $z_1 \bar w$. Similar considerations apply to the anti-collinear sector and the vertex at $z_2w$.

The leading power contribution in (\ref{leadingpo}) is expressed in terms of the non-perturbative matrix elements $f_1^q$, which was already mentioned in \eqref{delta}, as 
\begin{align}
\bra{P_2} O^q(0,0) \ket {P_1} = f_1^q(t,\mu_{hc}) \,  \bar N(P_2)\sbar n N(P_1).
\end{align}
We thus obtain a factorization theorem for the Feynman contribution to $H_q$  
\begin{align} \label{eq: Hq fact 2}
H_q^{\text{Feyn}}(x,\xi,t, \mu_{\rm UV}) = K_{qq}(x,\xi, t, \mu_{\rm UV},\mu_F) U(\mu_F,\mu_{hc})f_1^q(t,\mu_{hc}) + \f O(\lambda^{10}) ,
\end{align}
where the Fourier transform of the hard kernel  
\begin{align}
K_{qq}(x,\xi,t) = \bar n \cdot P\int \frac{dz}{2\pi} ds_1 ds_2 \, e^{i [(x-1) z -s_1  (1+\xi) + s_2 (1-\xi) ] \bar n \cdot P} \w K_{qq}(s_1,s_2, z),
\end{align} 
is  (\ref{kdef}) evaluated at $y=1$ and  given by  (\ref{Ky=1}). 
In \eqref{eq: Hq fact 2}, $\mu_{\rm UV}$ denotes the scale at which the GPD is renormalized, and $\mu_F$ denotes the factorization scale which separates the large scale $\sqrt{-t}$ from the hard-collinear scale $\mu_{hc}$. 
In practice, we can choose $\mu_{\rm UV}=\mu_F=\sqrt{-t}$ and resum the large double logarithms $\ln^2 (\sqrt{-t}/\mu_{hc})$ in the Sudakov factor $U$. It is worthwhile to mention that, at fixed coupling, 
\beq
 U(\sqrt{-t},\mu_{hc}) \sim \exp\left(-\frac{\alpha_sC_F}{2\pi} \ln^2\frac{-t}{\mu_{hc}^2}\right)  \sim \exp\left(-\frac{\alpha_sC_F}{8\pi} \ln^2\frac{-t}{\Lambda^2}\right). \label{u}
 \eeq 
The Sudakov suppression is significantly diminished  compared to the familiar formula $\sim e^{-\frac{\alpha_s C_F}{2\pi}\ln^2t}$ because of the $t$-dependence in $\mu^2_{hc} \sim \sqrt{-t}\Lambda$.

Following the same logic, it is straightforward to show that a similar factorization holds for the Feynman contribution to the gluon GPD $H_g$, with the sole non-perturbative input also being  $f_1^q$
\begin{align} \label{eq: Hg factorization}
H^{\rm Feyn}_g(x,\xi,t, \mu_{\rm UV}) = \sum_{q} K_{gq}(x,\xi, t, \mu_{\rm UV} , \mu_F)U(\mu_F,\mu_{hc}) f_1^{q}(t,\mu_{hc}) + \f O(\lambda^{10}).
\end{align}
Note that $K_{gq}$ starts at $\alpha_s(\sqrt{-t})$ for both $q = u,d$, so that the first term behaves as $\alpha_s/t^2$. The correction is naively power-suppressed  $\sim \lambda^{10} \sim 1/t^{5/2}$, but this term, being proportional to the matrix element of the operator $O^g$ in \eqref{eq: Og}, is not suppressed by a power of $\alpha_s(\sqrt{-t})$. Namely, this term leads to the contribution to $H_g$ 
\begin{align}
K_{gg}(x,\xi,t,\mu_{\rm UV}, \mu_F) \bra{P_2} O^g(0,0) \ket{P_1} \sim \frac{\alpha_s^0}{t^{5/2}},
\end{align}
 which is approximately of the same size as the term shown in \eqref{eq: Hg factorization} in the realistic parameter region where $\alpha_s(\sqrt{-t})/\pi \approx 1/\sqrt{-t}$.

The scale evolution of $K=(K_{qq},K_{gq})$ in $\mu_{\rm UV}$ is governed by the usual GPD evolution equations. Schematically, in terms of the evolution kernel $\mathbb H$,
\begin{align}
\frac{\partial}{\partial \ln \mu_{\rm UV}} \begin{pmatrix} K_{qq} \\ K_{gq}
\end{pmatrix} = \begin{pmatrix} \mathbb H_{qq} & \mathbb H_{qg} \\ \mathbb H_{gq} & \mathbb H_{gg} \end{pmatrix} \otimes \begin{pmatrix} K_{qq} \\ K_{gq}
\end{pmatrix}.
\end{align}
This is due to the fact that the UV properties are independent of the external states and therefore they are the same for the parton-level amplitudes $K$. 

The leading power SCET analysis thus shows that, in the pre-asymptotic regime, the quark GPD of the proton consists of two terms 
\begin{align}
H_q=H_q^{\text{hard}} + H_q^{\text{Feyn}} +{\cal O}\left(\frac{1}{t^{5/2}}\right) , \label{sum}
\end{align}
which both scale as $1/t^2$, but are factorized differently, \eqref{eq: hard scattering} and \eqref{eq: Hq fact 2}. An important caveat, first observed in \cite{Kivel:2012mf}, is that $H_q^{\rm hard}$ contains end-point divergences coming from the overlap region between soft and collinear modes. This implies that only the sum (\ref{sum}) 
is well-defined and consequently $H_q^{\rm Feyn}$ must also be divergent. However, the end-point divergences can be regulated using a suitable regulator, resulting in an additional scheme dependence that cancels between $H_q^{\text{hard}}$ and $H_q^{\text{Feyn}}$. In other words, $H_q^{\rm Feyn}$ is well-defined up to this scheme dependence. One may argue that since the divergence occurs at two-loops (meaning an order $\alpha_s(\sqrt{-t})^4$ term in $H_q^{\rm hard}$),    this dependence is negligible. Note that this is not guaranteed without actually constructing a definite regularization, which is beyond the scope of this paper. 

We note in passing  that the power counting argument is  significantly different for meson GPDs.   
In Appendix \ref{app: meson}, we consider meson external states and  show that  $H_{q/{\rm meson}}^{\rm hard} \sim \alpha_s\lambda^4 \sim \alpha_s/t$, while $H_{q/{\rm meson}}^{\rm Feyn} \sim \lambda^6 \sim 1/(-t)^{3/2}$. Importantly,   $H_{q/{\rm meson}}^{\rm hard}\propto\,   \alpha_s$, in contrast to  $H_{q/{\rm proton}}^{\rm hard}\propto\, \alpha_s^2$. This suggests that, in the pion GPD, the hard contribution becomes comparable to the Feynman contribution at a  smaller value of $|t|$, presumably already in the pre-asymptotic regime.

Returning to the proton case, the factorization formula \eqref{eq: Hq fact 2} has  immediate consequences  on the form factors. 
The function $f_1^q$ also appears in the Feynman contribution to the electromagnetic form factor $F_1$ as one can see  by integrating  \eqref{eq: Hq fact 2} over $x$ and summing over quark flavors    
\begin{align}
F_1^{\rm Feyn}(t) = C(t) U(t)\sum_q e_q f_1^q(\mu_{hc}) + \f O(\lambda^{10}), \label{remark}
\end{align}
where $C(t)$ is given by \eqref{eq: local hard} to one-loop. 
This formula was first derived in  \cite{Kivel:2010ns}. 
Together with \eqref{eq: Hq fact 2}, it leads to a surprising observation that the ratio of $H_q^{\rm Feyn}$ and $F_{1q}^{\rm Feyn}$ is purely perturbative up to power corrections 
\begin{align} \label{eq: ratio}
\frac{H_q^{\rm Feyn}}{F_{1q}^{\rm Feyn}} = \frac{K_{qq}(x,\xi,t)}{C(t)} + \f O(\lambda^{2}).
\end{align}
Furthermore, taking the second moment in $x$, we obtain a similar relation for the ratio between $F_1^{\rm Feyn}$ and the gravitational form factors  defined by 
\beq
\int_{-1}^1 dx x H_q(x,\xi,t,\mu)=A_q(t,\mu)+\xi^2D_q(t,\mu). \label{gffdef}
\eeq
In the present approximation we find $D_q(t)=0$ and 
\beq
A_q^{\rm Feyn}(t,\mu_{\rm UV})&=& \int_{-\xi}^1 dx x H_q^{\rm Feyn}(x,\xi,t,\mu_{\rm UV}) \\
&=& U(t)f_1^q(t)\left[1 + \frac{\alpha_s C_F}{4\pi} \left \{ - \ln^2  \frac{-t}{\mu^2_F}  + 3 \ln  \frac{-t}{\mu^2_F}  +\frac{8}{3}\ln \frac{-t}{\mu_{\rm UV}^2}- \frac{124}{9} + \frac{\pi^2}{6} \right \} +\f O(\alpha_s^2)\right] \nn && +{\cal O}(\lambda^{10}). \notag
\eeq 
 Again, the $\xi$-dependence cancels nontrivially between the DGLAP and ERBL contributions. 
We therefore obtain 
\begin{align}
\frac{A_q^{\rm Feyn}(t,\mu_{\rm UV})}{F_{1q}^{\rm Feyn}(t)} = 1 + \frac{\alpha_s C_F}{4\pi} \left ( \frac{8}{3} \ln \frac{-t}{\mu_{\rm UV}^2} - \frac{52}{9} \right ) + \f O\left ( \alpha_s^2 , \frac{1}{\sqrt{-t}}\right ) \approx 0.85, \label{08}
\end{align}
where $\alpha_s \equiv \alpha_s(\mu_{\rm UV})$. 
In the numerical estimate we have set $\mu^2_{\rm UV}=-t = 10 \,\text{GeV}^2$.
The leading corrections $\sim 1/\sqrt{-t}$ are the kinematical power corrections mentioned in (\ref{leadingpo}). The remaining corrections are of order $\alpha_s^2, \frac{\alpha_s}{\sqrt{-t}}$ and  $\frac{1}{t}$, coming from, for example, 
loop corrections to the $K$ kernel as well as the hard-scattering contribution \eqref{eq: hard scattering} ($\sim \alpha_s^2$), 
the contribution from $O^g$ ($\sim\alpha_s/\sqrt{-t}$), and the 
contribution from $O^{\bar q}$ $(\sim 1/t)$. 

Moreover, due to the factorization of the $x$-dependence, the ratio of any two moments of $H_q^{\rm Feyn}$ is perturbatively calculable up to power corrections. Specifically, let
\begin{align}
M_q^{(n)}(\xi, t) \equiv \int_{-\xi}^1 dx \, x^n H_q^{\rm Feyn}(x,\xi,t) = K_{qq}^{(n)}(\xi, t) f_1^q(t) \left [ 1 + \f O\left (\frac{1}{\sqrt{-t}} \right ) \right ].
\end{align}
One can check that polynomiality is satisfied for the one-loop kernel (\ref{Ky=1}). More precisely, for even $n$, $M_q^{(n)}$ as a function of $\xi$ is a polynomial in $\xi^2$ of degree $\frac{n}{2}$, while for odd $n$, $M_q^{(n)}$ is a polynomial in $\xi^2$ of degree $\frac{n-1}{2}$. 
The nonperturbative factor $f_1^q$ drops out in the ratio 
\begin{align}
\frac{M_q^{(n)}}{M_q^{(n')}} = \frac{K_{qq}^{(n)}}{K_{qq}^{(n')}} + \f O\left (\alpha_s^2, \frac{1}{\sqrt{-t}} \right ). \label{ratio2}
\end{align}
Importantly, even after setting $\mu^2_{\rm UV}=-t$, the moments $M_q^{(n)}$ have double logarithms  $M_q^{(n)} = 1 - \frac{\alpha_s C_F}{2\pi} \ln^2n +\cdots$ at large-$n$ coming from the $\left[\frac{\ln (1-x)}{1-x}\right]_+$ terms in (\ref{Ky=1}).  Since higher Mellin moments increasingly probe the large-$x$  region, this can  affect the  behavior of the GPD as $x\to 1$. Similarly to the resummation of threshold logarithms $\ln (1-x)$ in DIS \cite{Parisi:1979xd}, these logarithms should exponentiate $e^{- \frac{\alpha_s C_F}{2\pi} \ln^2n + ... }$.  Such a resummation ensures that $M_q^{(n)}$ goes to zero as $n \rightarrow \infty$, and accordingly, the GPD vanishes at $x=1$ as one might expect on physical grounds. 
We leave this problem for future work. (See however the next section.) For the moment, we note that the resummation does not affect \eqref{eq: ratio} (unless $x$ is too close to unity), {\color{red} nor}  low moments such as \eqref{08}.

We end this section with a few words of caution. First, although the formal power counting argument  justifies the inclusion of  higher order corrections in $\alpha_s$ to (\ref{08})  while neglecting power corrections in $\lambda^2\sim 1/\sqrt{-t}$,  the latter may not be numerically negligible in the pre-asymptotic regime.  We emphasize that the leading term `1'  in (\ref{08}) is a nontrivial and robust prediction, but already the order $\alpha_s/\pi$ term has  to compete with the leading  kinematical power correction $\sim \alpha_s^0/\sqrt{-t}$. In view of this, including higher order terms in $\alpha_s$ may not be justified for practical purposes.  
Note that this problem arises from the strict power counting argument resulting in the local $y$-dependence in the function $h_q(y)$  (\ref{delta}). In the next section, we suggest a (model-dependent) method to include the  kinematical power corrections by retaining a nontrivial $y$-dependence in  $h_q(y)$.

Second, it is somewhat questionable whether the SCET($c,\bar c,s$) power counting that we relied on holds if $\mu_{hc}$ is not a perturbative scale. Note that the formal power corrections from the SCET($C,\bar C,s$) $\rightarrow$ SCET($c,\bar c,s$) expansion are of the same size as for the QCD $\rightarrow$ SCET($C,\bar C,s$) expansion, namely $\f O(\mu_{hc}^2/t) = \f O(\Lambda^2/\mu_{hc}^2) = \f O(\Lambda/\sqrt{-t})$. However, since the matching coefficient $\f J_q$ is a power series in $\alpha_s(\mu_{hc})$, we cannot exclude the possibility that, when $\alpha_s(\mu_{hc})\sim 1$, large non-perturbative effects may  introduce a strong $t$ dependence and invalidate the formal power counting. Indeed, it has been suggested in a recent work \cite{Kivel:2024tgj} that the anti-quark contribution to the nucleon electromagnetic form factors may be numerically non-negligible around $|t|\approx 3$ GeV$^2$. {We note however that a similar issue arises in $B$-meson physics where the hard-collinear scale appears, see e.g. \cite{Beneke:2003pa}. Although $\sqrt{\Lambda m_b} \approx 1 \, \text{GeV}$,  the validity of the power counting argument is generally assumed. We therefore suppose that our analysis is on the same footing as such studies.

\section{Overlap representation}
\label{lc}

As we have seen, the  strict enforcement of the SCET power counting  rules leads to the leading-order, leading-power prediction $H_q^{\rm Feyn}(x,t) \approx \delta(1-x) f_1^q(t)$, which is unrealistic from a  phenomenological point of view.  In this section, we return to the more basic formula (\ref{fac2}) and model $h_q(y,t)$ using the so-called  overlap representation of GPDs \cite{Diehl:1998kh,Diehl:2000xz}  in the pre-asymptotic region. This  allows us to  establish a closer connection to the literature on the Feynman contribution   \cite{Drell:1969km,West:1970av}, and at the same time provides useful guidance for developing realistic models of GPDs at large-$t$. Formally, this approach attempts to include the kinematical power corrections in $1-y = \f O(\Lambda/\sqrt{-t})$, but it should be remarked that such a `hybrid' approach goes beyond the standard factorization framework in a physically motivated but heuristic way.

\subsection{DGLAP region}

The overlap representation of GPDs is a generalization of the parton model for the electromagnetic form factor. 
In this approach, GPDs are expressed by the convolution integral  (overlap) of light-cone wavefunctions in Fock space.  The quark GPD  in the DGLAP region $x>\xi$ reads   \cite{Diehl:2000xz}
\beq
H_q(x,\xi,t) &=& \sum_{N=3}^\infty \int dy d^2p_\perp\int \prod_{i=1}^{N-1} dx_i d^2k_{\perp i} \delta(x-y) \nn 
&&\times \Phi_N^*\left( \frac{y-\xi}{1-\xi}, p_\perp +\frac{1-y}{1-\xi}\frac{\Delta_\perp}{2};\frac{x_i}{1-\xi},k_{\perp i}-\frac{x_i}{1-\xi}\frac{\Delta_\perp}{2}\right) \nn 
&& \times\Phi_N \left(\frac{y+\xi}{1+\xi},p_\perp-\frac{1-y}{1+\xi}\frac{\Delta_\perp}{2};\frac{x_i}{1+\xi},k_{\perp i}+\frac{x_i}{1+\xi}\frac{\Delta_\perp}{2}\right). \label{over}
\eeq
$\Phi_N$ and $\Phi_N^*$ are the $N$-parton  light-cone wavefunctions (LCWFs) \cite{Lepage:1980fj,Brodsky:1997de} of the incoming and outgoing protons (not to be confused with the proton DAs $\Psi,\Psi^*$). $y$ and $p_\perp$ are the momentum fraction and transverse momentum of the active quark, 
 and $x_i$ and $k_{\perp i}$ are the corresponding variables for $N-1$ spectators. The  constraints $y+\sum_i^{N-1} x_i=1$, $p_\perp+\sum_i^{N-1} k_{\perp i}=0$ from momentum conservation are implicit.  As is clear from the use of light-cone wavefunctions, this model is intrinsically  tied to the infinite momentum frame where $P_1,P_2$ are both collinear. 
 The nontrivial arguments of $\Phi_N$ and $\Phi_N^*$ are a consequence of  Lorentz transformations (transverse boosts) between the `hadron-in' frames where   $P_{1\perp}=0$ and $P_{2\perp}=0$, respectively, and the `average' frame where $P_\perp=\frac{P_{1\perp}+P_{2\perp}}{2}=0$ (see \cite{Diehl:2000xz} for the details).  
Note that (\ref{over}) is not limited to large-$t$. In the forward limit $\xi=t=0$, it reduces to  the  unpolarized quark PDF $H_q(x,0,0) = q(x)$.

In principle, LCWFs can be rigorously defined within the framework of light-front quantization \cite{Brodsky:1997de}, and (\ref{over}) may be regarded as an exact, albeit formal representation. In practice, following \cite{Diehl:1998kh,Diehl:2000xz}, we use (\ref{over}) as a model of the Feynman contribution $H_q^{\rm Feyn}$ by removing the perturbative tail of the wavefunctions $\Phi_N(y,p_\perp)$. The nonperturbative part of $\Phi_N$ needs to be modeled, and naturally it has support only in the region $|p_\perp|\lesssim \Lambda$.  
This means that, at large momentum transfer $\Delta_\perp \sim \sqrt{-t}\gg \Lambda$, $y$ is forced to be close to unity
 \beq
 1-y\sim \frac{\Lambda}{\sqrt{-t}}\ll 1. \label{end}
 \eeq
 Due to the delta function $\delta(x-y)$, $H_q(x)$ becomes increasingly peaked in the large-$x$ region as $\sqrt{-t}$ is increased. At the same time, all spectator partons become soft $x_i\sim \frac{\Lambda}{\sqrt{-t}}\sim \lambda^2\ll 1$.  This is indeed the physical picture of the Feynman contribution.

\begin{figure}
\vspace{-40mm}
    \centering
\includegraphics[width=1\linewidth]{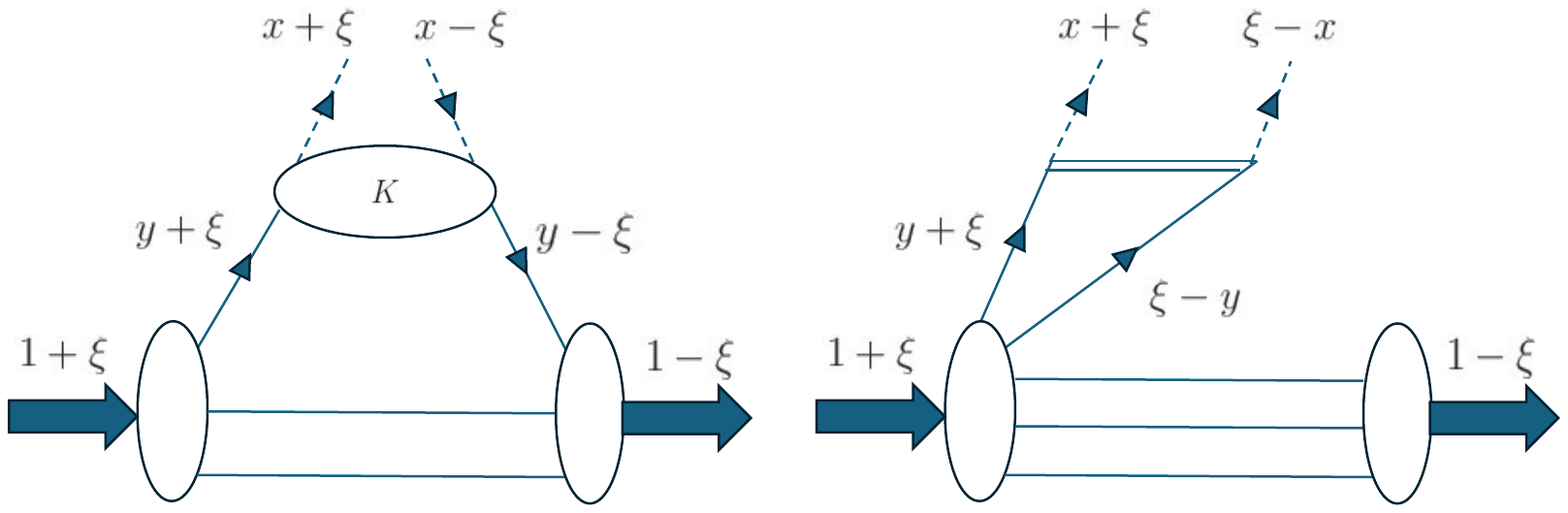}
\vspace{-35mm}
    \caption{Left: Proton GPD in the DGLAP region $\xi < x$. $y$ varies in the range $x\le y\le 1$. Right: Proton GPD in the ERBL region $x<\xi$ when $|t|\lesssim \Lambda^2_{\rm QCD}$. The diagram corresponds to $y<\xi$.   }
    \label{fig: GK model}
\end{figure}

 Our strategy is to incorporate the overlap representation into our SCET  analysis in Sections  \ref{proton new} and \ref{sec: SCET soft}. Let us  return to the general formula 
(\ref{fac2}). Instead of imposing the strict power counting argument in Sec. \ref{sec: SCET soft} on $h_q(y)$, which led to (\ref{delta}), we introduce a slight nonlocality in $y$ motivated by (\ref{end}).   This can be done straightforwardly (albeit somewhat  heuristically), by noticing that the delta function in (\ref{over}) is nothing but the quark-in-quark GPD at the tree level $K(x,y)\approx \delta(x-y)$. We thus identify the integrand of (\ref{over}) with $h_q(y)$ } and obtain the QCD-improved version of the overlap representation 
\begin{align}
H^{\rm Feyn}_q(x,\xi,t,\mu_{\rm UV}) =\sum_N\int_x^1 dy \int d^2p_\perp \prod_{i } dx_i d^2p_{\perp i} K_{\rm DGLAP}(x,y,\xi, t,\mu_{\rm UV},\mu_{F})U(\mu_{F},\mu_{hc})\Phi_N^* \Phi_N. \label{improve}
\end{align}
 The same Sudakov factor $U$ must be present\footnote{  Strictly speaking, when $y\neq 1$, the Lorentz transformations that bring the proton and the active quark to their respective Breit frames are different (see Appendix \ref{app A}). However, this does not affect the analysis in this section because the $y$-integral is practically limited to the range (\ref{end}) where the difference can be neglected. Besides, one can still impose $t_{\rm proton}=t_{\rm quark}$ even when $y\neq 1$, see Appendix \ref{app A}. Therefore the same Sudakov factor  can be used for  the quark and proton GPDs.    in order to resum large logarithms, and also to cancel the $\mu_F$-dependence of the kernel $K$. $\mu_{\rm hc}$ (by definition $\mu_{hc}\lesssim 1$ GeV in the pre-asymptotic regime) is here considered the defining scale of the model $\Phi_N\Phi^*_N$. } See Fig.~\ref{fig: GK model} (left) for a graphical representation of \eqref{improve}.  

To proceed, we keep only the leading Fock state $N=3$ in accordance with the suppression of  higher Fock states discussed in Sec. \ref{sec: SCET soft} and Appendices \ref{app: pc all orders} and \ref{app: meson},  and use the following Gaussian model 
\beq
\Phi_3\, \propto \,  (1-y)^\delta \exp\left(-\frac{p_\perp^2+k_{\perp 1}^2+k_{\perp 2}^2}{\Lambda^2}\right), \label{model}
\eeq
setting $\xi=0$ for simplicity. Due to the condition (\ref{end}), 
we are only concerned with the end-point behavior $y\to 1$ of $\Phi_3$  that  has been parametrized  in the form $(1-y)^\delta$. For example, if  $\Phi_3\, \propto\, x_1x_2y$ as $x_{1,2}\to 0$  with their ratio $x_1/x_2$ fixed,  then the condition $x_1+x_2+y=1$ implies $x_{1,2}\sim 1-y$, so that $\delta=2$.  More sophisticated models are available in the literature (e.g., \cite{Nair:2024fit}), but (\ref{model}) is sufficient to illustrate our point.\footnote{Another class of models, originally due to \cite{Brodsky:1981jv}, takes the form 
\beq
\Phi_3\propto \exp\left(-\frac{1}{\Lambda^2}\left(\frac{p_\perp^2}{y}+\frac{k_{\perp 1}^2}{x_1}+\frac{k_{\perp 2}^2}{x_2}\right)\right). 
\eeq
Substituting this into (\ref{improve}) and integrating over transverse momenta, we find 
\beq
H^{\rm Feyn}_q(x,t)\sim \int_x^1 dy dx_1dx_2\delta(1-y-x_1-x_2)x_1x_2y\exp\left(\frac{1-y}{2y}\frac{t}{\Lambda^2}\right).
\eeq
This means that the spectators have  longitudinal and transverse momenta  $x_{1,2}\sim 1-y\sim \frac{\Lambda^2}{-t}\sim \lambda^4$ and   $k_{\perp 1,2}\sim \sqrt{x_{1,2}}\Lambda \sim \frac{\Lambda^2}{\sqrt{-t}}=\lambda^4\sqrt{-t}$, respectively. These `ultrasoft' modes $p_{us}^\mu=\sqrt{-t}(\lambda^4,\lambda^4,\lambda^4)$ are unphysical since their virtuality is much softer than $\Lambda\sim \Lambda_{\rm QCD}$. See also a related critique \cite{Burkardt:2004bv} of this type of model.
} 
The $k_{\perp i}$ integrals reduce to an overall constant which we ignore.
In terms of $h_q$ in (\ref{fac2}), we find 
\beq
h_q(y,\xi = 0,t ) &\propto&  (1-y)^{2\delta} \int_0^{1-y} dx_1 \, \exp\left(\frac{t}{2\Lambda^2}((1-y)^2+x_1^2+(1-y-x_1)^2)\right) \nn 
&=& -\frac{\sqrt{\pi} M}{\sqrt{-t}}(1-y)^{2\delta}e^{\frac{3t(1-y)^2}{4\Lambda^2}} {\rm Erf}\left(-\frac{\sqrt{-t}}{2M}(1-y)\right) \nn 
&\approx & (1-y)^{2\delta+1}\exp\left(\frac{5t}{6\Lambda^2}(1-y)^2+{\cal O}(t^2(1-y)^4)\right) , 
\eeq
where Erf$(x)$ denotes the error function. 
The $y$-integral  gives, schematically, 
\beq
H_q^{\rm Feyn}(x)&\sim& \int_x^1 dy (1-y)^{2\delta+1} 
\exp\left(\frac{5t}{6\Lambda^2}(1-y)^2+\cdots\right)K_{\rm DGLAP}(x,y) \nn 
&=& (1-x)^{2\delta+1}\exp\left(\frac{5t}{6\Lambda^2}(1-x)^2+\cdots\right)+{\cal O}\left(\frac{\alpha_s}{\pi}\left(\frac{\Lambda^2}{-t}\right)^{\delta+1} \right). \label{est}
\eeq
The delta function in $K(x,y)\approx \delta(x-y)$ leads to a term that decays exponentially in $t$, whereas  the order $\alpha_s$ term in $K$ is a continuous function of $y$ and thus leads to a power-law behavior coming from the end-point of the $y$-integration
\beq
\int^1_{1-\frac{\Lambda}{\sqrt{-t}}}dy  (1-y)^{2\delta+1} \sim \left(\frac{\Lambda^2}{-t}\right)^{\delta+1}. \label{note}
\eeq
The region of the $y$-integral away from the end-point only gives an exponentially suppressed contribution $\sim e^{t/\Lambda^2}$.
Note that the leading order term also gives the same power-law  behavior after the $x$-integration (i.e., in form factors) 
\beq
\int^1_{1-\frac{\Lambda}{\sqrt{-t}}}dx  (1-x)^{2\delta+1} \sim \left(\frac{\Lambda^2}{-t}\right)^{\delta+1}. \label{note}
\eeq
However, at the GPD level, it is exponentially suppressed unless $x$ is  close to 1. This can be understood as the smeared version of the delta function $\delta(1-x)$ obtained in  the leading power SCET analysis   \eqref{eq: Hq fact 2}.  After this smearing,  (\ref{est}) actually vanishes at $x=1$, instead of diverging at $x=1$.

What should be the value of $\delta$? A common view \cite{Nesterenko:1983ef,Braun:2006hz,Braun:2001tj} is that the Feynman contribution to the form factor is a `higher twist correction' $F_1(t)\sim 1/t^3$ to the hard scattering  contribution $F_1(t)\sim \alpha_s^2/t^2$ in the pre-asymptotic regime. This implies $\delta=2$, which is reminiscent of the asymptotic behavior of the proton DA $\Psi \sim x_1x_2y\sim (1-y)^2$. 
However, the linear behavior $\delta=1$, suggested by a diagrammatic argument  \cite{Lepage:1980fj,Kivel:2012mf} and also by our SCET analysis in  Sec. \ref{sec: SCET soft}, is not excluded. Indeed, $\delta=1$ is consistent with the well-known dipole parametrization of the electromagnetic form factor $F_1(t)\sim 1/(1-t/M^2)^2$ with $M^2\approx 0.71$ GeV$^2$ \cite{Punjabi:2015bba}. It is also  natural to expect that the Feynman contributions in the pre-asymptotic and asymptotic regimes are smoothly connected such that the power behavior $F^{\rm Feyn}_1(t)\sim H^{\rm Feyn}_q(x,t)\sim 1/t^2$ is always the same, although the dependence on $\alpha_s$ may change.

\begin{figure}
    \centering
\includegraphics[width=0.4\linewidth]{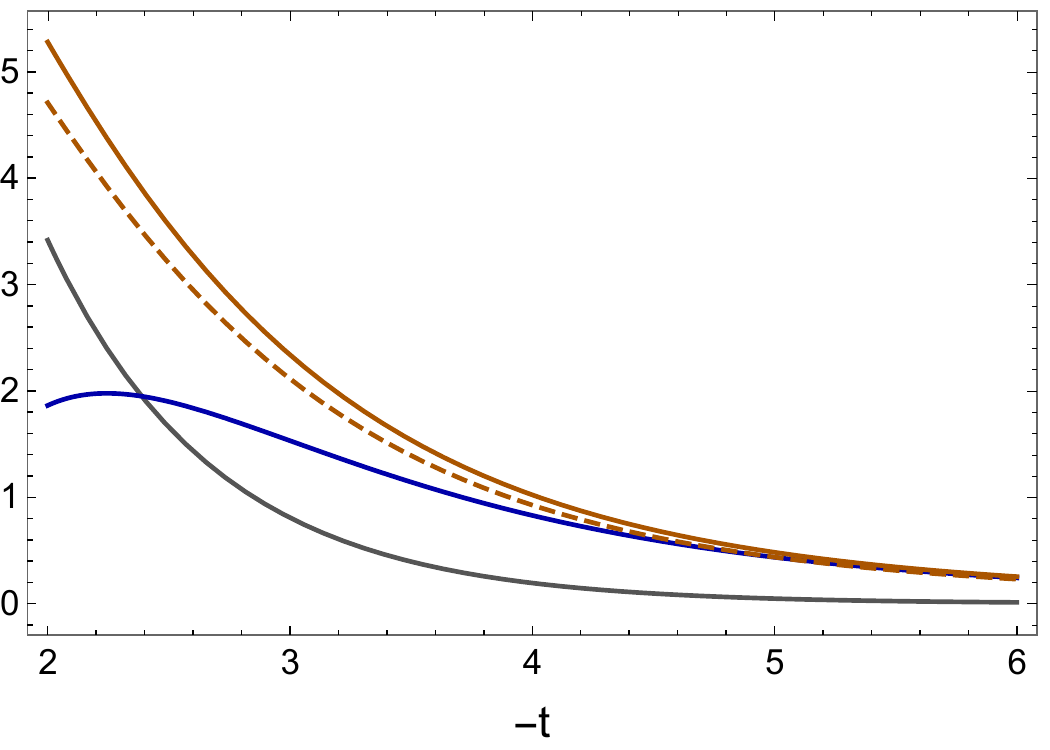} \qquad \includegraphics[width=0.41\linewidth]{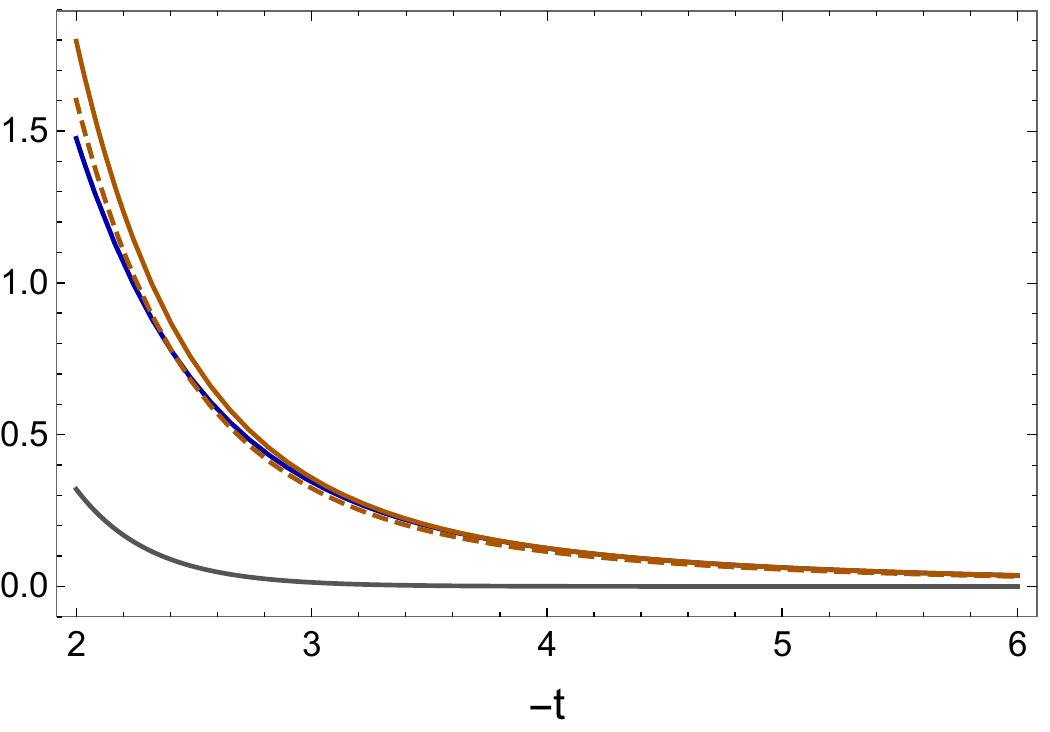}
\caption{ The Feynman contribution to the quark GPD $H^{\rm Feyn}_q$ at $x=0.6$ (left) and $x=0.4$ (right) with  $\xi=0$   as a function of $-t$ in units of GeV$^2$. (The $y$-axis is arbitrary.) Orange curve: Total $H_q^{\rm Feyn}$ without the Sudakov factor. The black and blue curves denote the  leading order $\alpha_s^0$ and the next-to-leading order $\alpha_s$ contributions, respectively.  The dashed curve denotes the total $H^{\rm Feyn}_q$ with the NNLL Sudakov factor $U$ included.}
    \label{fig: plot}
\end{figure}

In order to be more quantitative, we numerically evaluate (\ref{improve}) using (\ref{model}) and 
\beq
&&K_{\rm DGLAP}(x,y)= U(\sqrt{-t},\mu_{hc})\Biggl[\delta (x-y) \left(1+ \frac{\alpha_sC_F}{4\pi}  \biggl\{\frac{\pi^2}{6}-\ln^2\frac{x^2}{(1-x)^2} \biggr\}\right) \\
&& \quad +\frac{\alpha_sC_F}{2\pi} \left(\frac{(x^2+y^2)\ln (y^2)-(x-y)^2}{y^2}\left[\frac{\theta(y-x)}{y-x}\right]_+ -2\frac{x^2+y^2}{y^2}\left[\frac{\theta(y-x)\ln(y-x)}{y-x}\right]_+  \right) \Biggr]. \nonumber
\eeq
This is the RG-improved version of (\ref{d}) with the UV pole removed and $\mu_{\rm UV}^2=-t$. 
In Fig.~\ref{fig: plot}, we plot  $H_q^{\rm Feyn}$  at $x = 0.6$ (left) and $x=0.4$ (right) with $\xi = 0$, $\delta=1$, $\Lambda=0.3$ GeV as a function of $-t$. The normalization is arbitrary. 
The dashed curve includes the Sudakov factor, which we evaluated to next-to-next-to leading logarithmic (NNLL) accuracy  (i.e., using the three-loop cusp anomalous dimension and the four-loop beta function in the formulas in Section \ref{rgeq}). As expected, the order $\alpha_s$ term dominates over the leading order $\alpha_s^0$ term at large-$t$, and this tendency becomes more conspicuous as $x$ is lowered. We also see that the effect of the Sudakov factor is moderate, as noted in (\ref{u}). 

\subsection{ERBL region} 
\label{erblregion} 

We now turn to the overlap representation in the ERBL region $x<\xi$. Without going into the details, we schematically write the result of \cite{Diehl:2000xz} as 
\beq
H^{\rm Feyn}_q(x,\xi,t)\simeq \sum_{N=4} \int dy dp_\perp\int \prod_i^{N-1} dx_idp_{\perp i} \delta(x-y) \Phi^*_{N-1}\Phi_{N+1}. \label{rep}
\eeq
See Fig.~\ref{fig: GK model} (right).  The returning quark with momentum fraction $x-\xi<0$ in Fig.~\ref{fig: GK model} (left) has been reinterpreted as an outgoing antiquark with momentum fraction $\xi-x>0$. Thus the GPD in this region can be written as the superposition of  transition amplitudes  from  $(N+1)$-parton Fock states to $(N-1)$-parton Fock states. The lowest relevant Fock components $\Phi_5$ consist of  $qqqq\bar{q}$ states.  (\ref{rep}) represents processes in which a quark-anti-quark ($q\bar{q}$) pair with  total longitudinal momentum $(y+\xi)P^++(\xi-y)P^+=2\xi P^+$ and invariant mass $t$ is emitted from the incoming proton and gets absorbed by the operator $\bar{\psi}\sbar n \psi$.  
The remaining $N-1$ partons reassemble themselves to form the outgoing proton. 

However, the representation (\ref{rep}) is seldom used in practical calculations. For one thing, it is difficult to explicitly construct Fock states with five (anti-)quarks. For another, such a construction will in general violate  the polynomiality of GPD moments which requires an intricate relationship between the DGLAP and ERBL regions \cite{Chouika:2017dhe}. 
Regardless, we believe that the contribution (\ref{rep}) is exponentially suppressed  at large-$t$ because it is very  unlikely to find a $q\bar{q}$ pair with a large negative invariant mass $t$ in the proton wavefunction. We presume that this is  related  to the  leading term in (\ref{est}) in a way consistent with polynomiality. 
The dominant contribution in the ERBL region at $-t\gg \Lambda^2$ should take the  form 
\begin{align}
H^{\rm Feyn}_q(x,\xi,t,\mu_{\rm UV}) \approx  \int_\xi^1 dy \int dp_\perp \int\prod_{i }^{2} dx_i d^2p_{\perp i} K_{\rm ERBL}(x,y,\xi, t,\mu_{\rm UV},\mu_F)U(\mu_F,\mu_{hc})\Phi^*_3  \Phi_3(...,\mu_{hc}) , \label{erb}
\end{align}
 where the hard kernel $K_{\rm ERBL}$ is given by  (\ref{der}) to one-loop. At higher order, it will receive the same  Sudakov suppression $U$ as in the DGLAP region.  
Note that  the returning quark has a positive momentum fraction $y-\xi>0$. This is kinematically forbidden at the tree level $x=y<\xi$ but it becomes possible starting at one-loop. The $y$-integral in (\ref{erb}) will be  again dominated by the end-point region (\ref{end}). Therefore   the $t$-dependence in the ERBL region is parametrically the same as the order $\alpha_s$ term in (\ref{est}) in the DGLAP region.

It should be mentioned that the above discussion misses the so-called D-term \cite{Polyakov:1999gs}
\beq
H^D_q(x,\xi,t)=-E^D_q(x,\xi,t) = \theta(\xi-|x|) D_q(x/\xi,t). 
\eeq 
Since this is  part of the GPD $E_q$, its proper description  requires the inclusion of one unit of quark orbital angular momentum $L_z=\pm 1$ in $\Phi_3\Phi^*_3$  (cf., \cite{Tong:2022zax,Nair:2024fit}), which is beyond the scope of this work. 
However, since $D_q(x/\xi,t)$ is also part of $H_q$ in the ERBL region, the representation (\ref{erb}) is still valid aside from the choice of the wavefunction. Therefore, $D_q(x/\xi,t)$ at large-$t$ is again dominated by the order $\alpha_s$ contribution and receives  the Sudakov suppression. On the other hand, the wavefunction does affect the power-law behavior. Since the orbital excitation of Fock states typically causes an extra power of $1/t$  \cite{Tong:2022zax,Nair:2024fit}, we estimate that
\beq
D^{\rm Feyn}_q(x/\xi,t,\mu) \sim U(\mu,\mu_{hc}) \frac{\alpha_s}{\pi} \left(\frac{\Lambda^2}{-t}\right)^{\delta+2}.  \qquad (-t \gg \Lambda^2) \label{dqx}
\eeq

\subsection{Gravitational form factors}
Finally, we briefly discuss the implications of the above results for the gravitational form factors (GFFs) that are contained in the second moment of the GPD (\ref{gffdef}). 
As we have already noted below (\ref{note}),  the $x$-integral converts the exponentially suppressed  term in (\ref{est}) into a power-law contribution. Since this is order $\alpha_s^0$, it will dominate over the integral of the next-to-leading order  $\alpha_s$ term in  (\ref{est}). 
 We thus find 
\beq
F^{\rm Feyn}_{1q}(t,\mu)\sim 
A^{\rm Feyn}_q(t,\mu)\sim U(\mu,\mu_{hc})\left(\frac{\Lambda^2}{-t}\right)^{\delta+1}. 
\eeq
As for the $D$-type GFF, we do not see this interesting interplay between the order  $\alpha_s^0$ and  $\alpha_s$ contributions since the $x$-integral $-\xi<x<\xi$ does not include the  region $x\sim 1$.  From (\ref{dqx}), we find
\beq
D^{\rm Feyn}_q(t,\mu) \sim U(\mu,\mu_{hc}) \frac{\alpha_s(\mu)}{\pi} \left(\frac{\Lambda^2}{-t}\right)^{\delta+2}. \label{appa}
\eeq

The above estimates should be contrasted with the hard scattering  contribution  \cite{Tanaka:2018wea,Tong:2022zax} 
\beq
F^{\rm hard}_q(t,\mu)\sim A^{\rm hard}_q(t,\mu)\sim \left(\frac{\alpha_s(\mu)}{\pi}\right)^2 \left(\frac{\Lambda^2}{-t}\right)^2, \qquad D_q^{\rm hard}(t,\mu) \sim \left(\frac{\alpha_s(\mu)}{\pi}\right)^2 \left(\frac{\Lambda^2}{-t}\right)^3. \label{lt}
\eeq
If $\delta=2$, as is often assumed, the Feynman contribution  dominates over the leading twist contribution in $F_1(t),A_q(t)$  up to $-t\lesssim \left(\frac{\pi}{\alpha_s}\right)^2\Lambda^2\sim 100\Lambda^2$. 
In addition, the extra factor of $x$ in the integrand (\ref{gffdef}) in the gravitational case (as opposed to the electromagnetic case)   further emphasizes the Feynman contribution since $H_q^{\rm Feyn}$  tends to be supported in the large-$x$ region at large-$t$. 
Interestingly, in $D_q(t)$, the two contributions become comparable at a much lower scale $-t \sim \frac{\pi}{\alpha_s}\Lambda^2\sim 10\Lambda^2$. 
On the other hand, if $\delta=1$, which we deem a viable possibility, the Feynman contribution always dominates in the pre-asymptotic regime, although eventually it will be suppressed in the asymptotic limit $|t|\to \infty$. 
 
\section{Summary and conclusions}

In this paper we have studied the Feynman contribution to the quark GPD and related form factors  in the `pre-asymptotic' regime loosely defined by (\ref{re1}).   
 Our main findings can be summarized as follows.

\begin{itemize}
\item GPDs develop large double logarithms at large-$t$ and they can be resummed into the Sudakov form  factor. This has been shown first at the quark level and then at the hadronic level \eqref{eq: Hq fact 2}. Our result generalizes the recent one-loop analysis in  \cite{Bhattacharya:2023wvy} to all orders and enables a consistent resummation of all the large logarithms $\ln Q^2$ and $\ln^2 t$ (and subleading logarithms in $t$) in two-scale exclusive processes such as DVCS in the kinematical regime $\Lambda_{\rm QCD}^2\ll |t| \ll Q^2$. We emphasize that this issue has not been adequately addressed in the GPD literature.   

\item
We have performed a SCET-based power counting analysis in the pre-asymptotic regime and found that the ratio $H_q^{\rm Feyn}/F_{1q}^{\rm Feyn} = \delta(1-x) + \f O(\alpha_s(\sqrt{-t}))$ \eqref{eq: ratio} is a perturbative quantity up to power corrections. Similar relations hold for the ratio of moments of GPDs including the gravitational form factor \eqref{08}. However, it remains to be seen whether non-perturbative effects, due to the hard-collinear scale $\mu_{hc}^2=\Lambda^2 \sqrt{-t}$ being a non-perturbative scale, spoil this estimate by introducing a strong $t$-dependence and make the power corrections relatively important.

\item 
We have obtained  the following parametric $t$- and $\alpha_s$-dependencies of the proton  GPDs (when $x$ is not too close to unity) and GFFs 
\beq
H_q(x,\xi,t) &\sim& U\frac{\alpha_s}{\pi}\left(\frac{\Lambda^2}{-t}\right)^{\delta+1} + \left(\frac{\alpha_s}{\pi}\right)^2 \left(\frac{\Lambda^2}{-t}\right)^2, \nn 
D_q(x/\xi,t) &\sim& U\frac{\alpha_s}{\pi}\left(\frac{\Lambda^2}{-t}\right)^{\delta+2} + \left(\frac{\alpha_s}{\pi}\right)^2 \left(\frac{\Lambda^2}{-t}\right)^3,\nn 
A_q(t) &\sim& U\left(\frac{\Lambda^2}{-t}\right)^{\delta+1} + \left(\frac{\alpha_s}{\pi}\right)^2 \left(\frac{\Lambda^2}{-t}\right)^2, \nn 
 D_q(t) &\sim& U\frac{\alpha_s}{\pi}\left(\frac{\Lambda^2}{-t}\right)^{\delta+2} + \left(\frac{\alpha_s}{\pi}\right)^2 \left(\frac{\Lambda^2}{-t}\right)^3, 
\eeq 
 where $\alpha_s=\alpha_s(\sqrt{-t})$ and $U$ is the abbreviation of the Sudakov factor  (\ref{cu}) with $\mu_F\sim \sqrt{-t}$ and $\mu_{0}\sim \mu_{hc}= (-t\Lambda^2)^{\frac{1}{4}}$.
 The first term is our new result for the Feynman contribution  and the second term is the known hard scattering  contribution \cite{Hoodbhoy:2003uu,Tong:2022zax}. The parameter $\delta$ dictates the end-point behavior of the light-cone wavefunction $\Phi_3$ (\ref{model}). Our SCET analysis in Sec. \ref{sec: SCET soft} (see also Appendix \ref{app: pc all orders}) suggests  $\delta=1$. 
Since the suppression due to the Sudakov factor $U$ is limited, the Feynman contribution always dominates over the hard scattering contribution, especially in the $A_q$ GFF. The dominance is relatively less pronounced in $H_q(x,t)$, $D_q(x,t)$ and $D_q(t)$ because of the factor $\frac{\alpha_s}{\pi}\sim 0.1$  but is still there in a certain window of $t$ depending on the parameter $\delta$. Therefore, GPDs and GFFs in the phenomenologically relevant region of $|t|$ are dominated by the Feynman contribution.  To go beyond these parametric estimates, one needs to specify the light-cone wavefunction, see e.g., \cite{Nair:2024fit}. 

\end{itemize}

The behavior of GPDs and GFFs at moderately large-$t$ is important for phenomenology such as wide-angle Compton scattering \cite{Radyushkin:1998rt,Huang:2001ej} and the  near-threshold production of heavy vector mesons such as $\phi$ and $J/\psi$   in electron scattering, see the recent developments  \cite{Hatta:2025vhs,Guo:2025jiz,Hatta:2025ryj} and references therein.  Via crossing, our discussion  is  relevant to the study of the  generalized distribution amplitudes in the timelike region $t\gg \Lambda^2_{\rm QCD}$ \cite{Diehl:1999ek,Song:2025zwl,Gousset:1994yh}. The large-$t$ behavior of GFFs is also of interest from  general relativity perspective \cite{Hatta:2023fqc,Coriano:2024qbr,Dumitru:2025gzc}.  We hope  our results are useful for better parametrizations and extractions of GPDs and GFFs in these studies.

\section*{Acknowledgments}
We are grateful to Vladimir Braun, Markus Diehl, Nikolay Kivel, Anatoly Radyushkin, Robert Szafron and Feng Yuan for useful discussions. 
We  were supported by the U.S. Department
of Energy under Contract No. DE-SC0012704, and also by LDRD funds from Brookhaven Science Associates.

\appendix

\section{Lorentz transformations} \label{app A}

In this Appendix, we construct explicit Lorentz transformations to arrive at the frame (\ref{breit}) starting from the `usual' frame 
\beq
p^\mu_1 = \left((y+\xi)P^+,\frac{a^2\Delta_\perp^2}{(y+\xi)P^+}, a\vec{\Delta}_\perp\right),\qquad 
p_2^\mu = \left((y-\xi)P^+,\frac{b^2\Delta_\perp^2}{(y-\xi)P^+}, b\vec{\Delta}_\perp\right),
\label{pp_2}
\eeq
where $p_1^2=p_2^2=0$ and we used the notation $p^\mu =(p^+,p^-,\vec{p}_\perp)$. We set 
\beq
a=-\frac{\xi+y}{2\xi} +\frac{1}{2\xi}\sqrt{\frac{y^2-\xi^2}{1-\xi^2}}, \qquad b=\frac{\xi-y}{2\xi} +\frac{1}{2\xi}\sqrt{\frac{y^2-\xi^2}{1-\xi^2}}, 
\eeq
such that the momentum transfer at the parton level 
\beq
\Delta^\mu=p_2^\mu - p^\mu_1 = \left(-2\xi P^+, \frac{\xi \vec{\Delta}_\perp^2}{2(1-\xi^2)P^+},\vec{\Delta}_\perp\right), \qquad  t=(p_2-p_1)^2=\frac{-\vec{\Delta}_\perp^2}{1-\xi^2},
\eeq
is the same as that  for the proton external states $(p_2-p_1)^2=(P_2-P_1)^2$ (neglecting the proton mass). The latter is obtained by setting $y=1$ where $b=-a=\frac{1}{2}$.

First, perform a transverse boost to  eliminate $\vec{\Delta}_\perp$ in $\Delta^\mu$  
\beq
\Delta^\mu \to \left(-2\xi P^+,\frac{\vec{\Delta}^2_\perp}{2\xi (1-\xi^2)P^+}, 0_\perp\right).
\eeq
Under this boost, 
\beq
p^\mu_1 &\to& \left((y+\xi)P^+,\frac{(y-\xi)\vec{\Delta}_\perp^2}{4\xi^2(1-\xi^2)P^+}, \frac{1}{2\xi}\sqrt{\frac{y^2-\xi^2}{1-\xi^2}}\vec{\Delta}_\perp\right),\nn
p_2^\mu &\to& \left((y-\xi)P^+,\frac{(y+\xi)\vec{\Delta}_\perp^2}{4\xi^2(1-\xi^2)P^+}, \frac{1}{2\xi}\sqrt{\frac{y^2-\xi^2}{1-\xi^2}}\vec{\Delta}_\perp\right)
\eeq
Next, perform a longitudinal Lorentz boost with the boost factor $\lambda = \frac{\sqrt{-t}}{2\xi P^+}$
\beq
\Delta^\mu \to 
\sqrt{-t}(-1,1,0_\perp) .
 \eeq
 Perform the same boost to $p_1,p_2$ 
 \beq
p^\mu_1 \to \frac{\sqrt{-t}}{2\xi}\left(y+\xi,y-\xi, \sqrt{y^2-\xi^2}\hat{\Delta}_\perp\right), \qquad  
p_2^\mu \to \frac{\sqrt{-t}}{2\xi}\left(y-\xi,y+\xi, \sqrt{y^2-\xi^2}\hat{\Delta}_\perp\right),
\eeq
where $\hat{\Delta}_\perp = \vec{\Delta}_\perp/|\vec{\Delta}_\perp|$. 
Switching now to the Minkowski notation  $p^\mu=(p^0,p^1,p^2,p^3)$, we write   
\beq
p^\mu_1 =\frac{\sqrt{-t}}{2\xi} \left( y, \sqrt{y^2-\xi^2},0, \xi\right), \qquad p^\mu_2 =\frac{ \sqrt{-t}}{2\xi}\left( y, \sqrt{y^2-\xi^2},0, -\xi\right),
\eeq
where, without loss of generality, we have  chosen $\vec{\Delta}_\perp$ to be in the $1$-direction.  
Boost along the $1$-direction with velocity $v=-\frac{1}{y}\sqrt{y^2-\xi^2}$, $\gamma=\frac{1}{\sqrt{1-v^2}} = \frac{y}{\xi}$
\begin{align}
p_1^\mu \to \frac{\sqrt{-t}}{2} \left(1,0,0,1\right) \equiv \frac{\sqrt{-t}}{2}\sqrt{\frac{y+\xi}{y-\xi}}w^\mu , \quad p_2^\mu \to \frac{\sqrt{-t}}{2}(1,0,0,-1) \equiv \frac{\sqrt{-t}}{2}\sqrt{\frac{y-\xi}{y+\xi}}\bar{w}^\mu.
\end{align}
This is the desired frame spanned by  the new light-cone vectors  $w^2=\bar{w}^2=0$,  $w\cdot \bar{w}=2$. 
 Finally, we perform the same set of transformations to $\bar n^\mu$ and find 
\beq
\bar  n^\mu=(0,2,0_\perp)  \to  \sqrt{\frac{y^2-\xi^2}{-t}}P^+\left( w^\mu + \bar{w}^\mu - 2\hat{\Delta}_\perp^\mu \right),
\eeq
where $\hat{\Delta}_\perp^\mu \equiv  (0,0, \frac{\vec{\Delta}_\perp}{|\vec{\Delta}_\perp|})$. Note that $\bar n^\mu$ is defined up to an overall factor since the rescaling $\bar n^\mu \to c\bar n^\mu$ can be absorbed by $z\to z/c$ in (\ref{def}). Using this freedom, we can finally write 
\beq
\bar n^\mu \to w^\mu + \bar{w}^\mu +  n_\perp^\mu, \qquad n_\perp^\mu = - 2\hat{\Delta}_\perp^\mu .
\eeq

\section{Power estimates for the hard-collinear matching kernels to all orders} \label{app: pc all orders}

\begin{figure}
    \centering
    \includegraphics[width=1\linewidth]{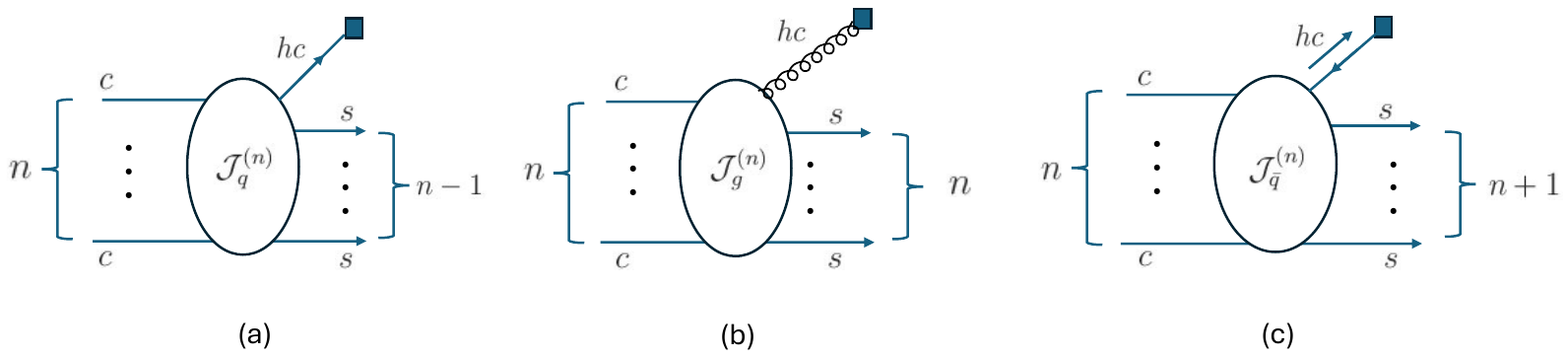}
    \vspace{-6.5cm}
    \caption{ Amplitudes for matching kernels with generic number $n$ of incoming collinear quarks for operators (a) $\xi_{hc}$  (b) $A_{hc\perp}$  (c) $\bar \xi_{hc}$
    }
    \label{fig: hcmatching general}
\end{figure}

In this appendix, we generalize our tree-level analysis in Sec.~\ref{sec: SCET soft} (see Fig.~\ref{fig: hcmatching}) to all orders. We use an argument based on dimensional analysis and Lorentz covariance, which is similar to how the graphical power counting rules are derived in the traditional pQCD approach, see e.g. section 5.8 of \cite{Collins:2011zzd}.

Start with the quark diagram Fig.~\ref{fig: hcmatching general}(a) with $n$ being the number of incoming collinear quarks. 
First of all, a generic scattering amplitude has dimension $4-\#(\text{external legs})$. To obtain the dimension of $\f J_q^{(n)}$ we need to subtract $\frac{1}{2}(2n-1)$ to take into account the fact  that the external collinear and soft quarks are amputated. Furthermore we need to subtract $\frac{3}{2}$ because the hard-collinear quark is not amputated. This gives
\begin{align}
\text{dim} \, \f J_q^{(n)} = 4 - 2n - \frac{3}{2} (2n-1) - \frac{3}{2} = -3n + 3. 
\end{align}
Similar considerations show that
\begin{align}
\text{dim} \, \f J_g^{(n)} &= 1-3n,
\\
\text{dim} \, \f J_{\bar q}^{(n)} &= -3n ,
\end{align}
for the gluon and anti-quark diagrams Fig.~\ref{fig: hcmatching general} (b) and (c), respectively.

Consider an arbitrary quark line $\Gamma_{\alpha \beta}$ (here $\alpha, \beta$ are Dirac indices ) in $\f J_q^{(n)}$, which must start at an external collinear quark line and end at an external soft quark line or at the hard-collinear operator (represented by a box in the diagram). Then $\Gamma$ appears in the spinor products 
\begin{align} \label{eq: spinor products}
\bar q_s \Gamma \xi_c \qquad \text{ or } \qquad (...) \s{\bar w} \s p_{hc} \Gamma \xi_c,
\end{align}
where $p_{hc} \sim (1, \lambda^2 , \lambda^2)$ is momentum that flows into the hard-collinear operator $\xi_{hc}$. Note that $p_{hc}$ must be a linear combination of collinear and soft momenta, so that $p_{hc\perp} \sim \lambda^2$.
Since helicity is conserved in the hard-collinear scattering (masses are set to zero) $\Gamma$ can be reduced to either $\gamma^{\mu}$ or $\gamma^{\mu} \gamma_5$. Since $\gamma_5$ does not play a role in the following argument, it will be ignored below. 
We can write the amplitude, including the external operators, as
\begin{align}
A_{\mu_1 ... \mu_n} [(...) \s{\bar w} \s p_{hc} \gamma^{\mu_1} \xi_c ] [\bar q_s \gamma^{\mu_2} \xi_c] \, ... \, [ \bar q_s \gamma^{\mu_n} \xi_c ]. 
\end{align}
The quantity $A_{\mu_1 ... \mu_n}$  transforms in a Lorentz covariant way, so we can determine its scaling in the rest frame \cite{Collins:2011zzd}, where its scaling dimension is given by its mass dimension
\begin{align} \label{eq: rest frame scaling}
A_{\mu_1 ... \mu_n}|_{\text{rest}} \sim \lambda^{\text{dim}\, A}.
\end{align}
This is implied by noting that each momentum in $\f J_q^{(n)}$ is hard-collinear by definition. Furthermore, transverse components of external momenta are set to zero in $\f J_q^{(n)}$, while hard-collinear loop momenta scale as $k_{hc} \sim (1, \lambda^2, \lambda)$. In the rest frame of $\f J_q^{(n)}$ all (non-zero) momentum components thus scale in the same way, i.e. $\lambda$, which implies \eqref{eq: rest frame scaling}.

This scaling is retained in the original boosted frame for the $\perp$ components of $A$. Then, for each index $\mu_j$ that is contracted with $\bar w^{\mu_j}$ the rest frame scaling is modified by $\lambda^{-1}$ and for each $\mu_j$ that is contracted with $w^{\mu_j}$ the rest frame scaling is modified by $\lambda^1$. These properties can now be used to determine the $\lambda$-scaling. For odd $n$, we may pick $\mu_2,...,\mu_n = \perp$ and $\s p_{hc} \gamma^{\mu_1} \rightarrow \bar w \cdot p_{hc} w^{\mu_1} \, \frac{\s w {\sbar w}}{4}$. Since the $p_{hc}$ propagator is contained in $\f J_q^{(n)}$, picking the large component $\s p_{hc} \rightarrow \bar w \cdot p_{hc} \frac{\s w}{2}$ gives an enhancement by $\lambda^{-1}$ relative to the rest frame scaling. This enhancement is compensated by contracting the $\mu_1$ index with $w^{\mu_1}$. Hence
\begin{align}
\f J_q^{(n)} \sim A_{\mu_1 \perp ... \perp} \bar w \cdot p_{hc} w^{\mu_1} \sim \lambda^{\text{dim}\, \f J_q^{(n)}} = \lambda^{-3(n-1)} \stackrel{n=3}{=} \lambda^{-6}. \qquad (n \text{ odd})
\end{align}
Note that $n$ being odd is important because in $\f J_q^{(n)}$ transverse components of external momenta should be set to zero and only rank $2$ tensors are available in two transverse dimensions, namely $g_{\perp}^{\mu \nu}$ and $\varepsilon_{\perp}^{\mu \nu}$. Thus, for even $n$ we need to pick one of the $\mu_2 ,... , \mu_n$ to be contracted with $w$, giving an additional power of suppression. Thus
\begin{align}
\f J_q^{(n)} \sim \lambda^{\text{dim}\, \f J_q^{(n)} + 1} = \lambda^{-3n+4} \stackrel{n=2}{=} \lambda^{-2}. \qquad (n \text{ even})
\end{align}
This proves $\f J_{q}' \sim \lambda^{-2}$ (see \eqref{eq: xihc xic}) to all orders.

Next, consider $\f J_g^{(n)}$ in Fig. \ref{fig: hcmatching general}(b). The complete amplitude including external operators can be written as
\begin{align}
A_{\mu_0 \mu_1 ... \mu_n} \, [\bar q_s \gamma^{\mu_1} \xi_c] ... [\bar q_s \gamma^{\mu_n} \xi_c],
\end{align}
where $\mu_0$ is the transverse index that is contracted with the external hard-collinear gluon operator. For odd $n$ all components of $A$ can be choosen to be transverse, so
\begin{align}
\f J_g^{(n)} \sim \lambda^{\text{dim}\,\f J_g^{(n)}} = \lambda^{1-3n} \stackrel{n=3}{=} \lambda^{-8}, \qquad (n \text{ odd})
\end{align}
while for even $n$ one index has to be contracted with $w$, so
\begin{align}
\f J_g^{(n)} \sim \lambda^{\text{dim}\,\f J_g^{(n)} + 1} = \lambda^{2-3n}. \qquad (n \text{ even})
\end{align}
Finally, consider $\f J_{\bar q}^{(n)}$. The complete amplitude including external operators can be written as
\begin{align}
A_{\mu_0 \mu_1 ... \mu_n} \, [\bar q_s \gamma^{\mu_0} \s p_{hc} \s{\bar w} (...)] [\bar q_s \gamma^{\mu_1} \xi_c] ... [\bar q_s \gamma^{\mu_n} \xi_c].
\end{align}
For odd $n$, the minimum power of $\lambda$ can be obtained by choosing all $\mu_0,...,\mu_n = \perp$. Since we may pick $\s p_{hc} \rightarrow \bar w \cdot p_{hc} \frac{\s w}{2}$ we get an enhancement by $\lambda^{-1}$ relative to the rest frame scaling:  
\begin{align}
\f J_{\bar q}^{(n)} &\sim \lambda^{\text{dim}\,\f J_{\bar q}^{(n)} - 1} =  \lambda^{-3n-1} \stackrel{n=3}{=} \lambda^{-10} \qquad (n \text{ odd}).
\end{align}
For even $n$, the minimum power of $\lambda$ is obtained by picking $\mu_1,...,\mu_n = \perp$ and $\gamma^{\mu_0} \s p_{hc} = w^{\mu_0} \bar w \cdot p_{hc} \frac{\s{\bar w} \s w}{4}$. Thus
\begin{align}
\f J_{\bar q}^{(n)} &\sim \lambda^{\text{dim}\,\f J_{\bar q}^{(n)}} = \lambda^{-3n}. \qquad (n \text{ even})
\end{align}
Note that the above arguments can, with some straightforward modifications, also be used to infer the power counting for higher Fock states to the Feynman contribution. Say, for example, we want to consider the matching
\begin{align}
\xi_{C} \ket {P_1} \rightarrow \w{\f J}_q \, \bar q_s \bar q_s \xi_c \xi_c \xi_c \bar \xi_c \xi_c \ket{P_1}.
\end{align}
The leading contribution can be written as
\begin{align}
A_{\mu_1 \mu_2 \mu_3 \mu_4} [(...) \s{\bar w} \s p_{hc} \gamma^{\mu_1} \xi_c ] [\bar \xi_c \gamma^{\mu_2} \xi_c] [\bar q_s \gamma^{\mu_3} \xi_c] [ \bar q_s \gamma^{\mu_4} \xi_c ]. 
\end{align}
Since we are forced to pick $\s p_{hc} \gamma^{\mu_1} = w^{\mu_1} \bar w \cdot p_{hc} \frac{\s w \s{\bar w}}{4}$ and $\gamma^{\mu_2} = w^{\mu_2} \frac{\s{\bar w}}{2}$, we obtain two $2-1 = 1$ power of suppression relative to the rest frame scaling, i.e. $\w{\f J}_q \sim \lambda^{\text{dim}\,\f J_q^{(4)} + 1} = \lambda^{-8}$. More generally, replacing an outgoing soft quark with an incoming collinear anti-quark results in power suppression by $\lambda^1$. We can then imagine a scenario where we have the higher $qqq\bar q q$ Fock state for the incoming collinear proton and the valence $qqq$ Fock state for the outgoing anti-collinear proton. The power estimate for this contribution to the form factor is then
\begin{align}
\underbrace{ \lambda^{-8} }_{\sim \w {\f J_q}} \times \underbrace{ \lambda^{-6} }_{\sim \f J_q} \times \underbrace{  (\lambda^6)^2 }_{\sim (q_s \bar q_s)^2} \times \underbrace{ \lambda^4 }_{\sim \bra{P_2} \xi_{\bar c} \xi_{\bar c} \xi_{\bar c} } \times \underbrace{ \lambda^8 }_{\sim \xi_c \xi_c \xi_c \bar \xi_c \xi_c \ket{P_1}} = \lambda^{10}.
\end{align}

\section{SCET power counting for meson states}

\label{app: meson}

Our results for the Feynman contribution to the proton GPD in Sec. \ref{sec: SCET soft} can be readily generalized to meson states.  The power counting  formulas derived in Appendix \ref{app: pc all orders} immediately imply the corresponding power estimates for the meson (in particular, pion) matrix elements $\bra{P_2} O^{q,g,\bar q} \ket {P_1}$. The hard scattering contribution is given schematically by the matrix element
\begin{align}
\bra{P_2} \bar \xi_{\bar c} \xi_{\bar c} \bar \xi_c \xi_c \ket{P_1} \sim \lambda^4  \sim \frac{1}{t},
\end{align}
which is the leading power. The Feynman contribution due to the quark involves the matching
\begin{align}
\xi_{C} \ket {P_1} \rightarrow \widetilde{\f J}_q q_s \bar \xi_c \xi_c \ket{P_1},
\end{align}
where it is easy to see that $\widetilde{\f J}_q \sim \f J_q^{(2)} \sim \lambda^{-2}$, so we have
\begin{align}
\bra{P_2} O^q \ket{P_1} \sim (\lambda^{-2})^2 \times (\lambda^6)^1 \times (\lambda^2)^2 = \lambda^6 \sim \frac{1}{t^{3/2}}. 
\end{align}
The first factor is $(\f J_q^{(2)})^2 \sim (\lambda^{-2})^2$, the second factor is $q_s \bar q_s \sim \lambda^6$ and the third factor is $(\xi_c \xi_c \ket{P_1})^2 \sim (\lambda^2)^2$. It is easy to see that the anti-quark contribution is now of the same order as the quark contribution
\begin{align}
\bra{P_2} O^{\bar q} \ket{P_1} \sim \lambda^6,
\end{align}
as expected. This follows from the  matching
\begin{align}
\bar \xi_{C} \ket{P_1} \rightarrow \widetilde{\f J}_{\bar q} \bar q_s \bar \xi_c \xi_c \ket{P_1},
\end{align}
with $\widetilde{\f J}_{\bar q}\sim \f J_q^{(2)} \sim \lambda^{-2}$. The gluon contribution has the matching
\begin{align}
A_{C\perp} \ket{P_1} \rightarrow \widetilde{\f J}_g \bar q_s q_s \bar \xi_c \xi_c \ket{P_1},
\end{align}
with $\widetilde{\f J}_g \sim \f J_g^{(2)} \sim \lambda^{-4}$. Therefore, 
\begin{align}
\bra{P_2} O^g \ket{P_1} \sim (\lambda^{-4})^2 \times (\lambda^6)^2 \times (\lambda^2)^2 = \lambda^8.
\end{align}

\bibliographystyle{apsrev}

\bibliography{ref}

\end{document}